\pdfoutput=1

\documentclass[a4paper,fleqn]{cas-dc}

\usepackage{cas-common}
\usepackage{graphics}
\usepackage{graphicx}

\usepackage{hyperref}
\usepackage{threeparttable}
\usepackage{booktabs}
\usepackage{multirow}

\usepackage{multicol}
\usepackage{makecell}
\usepackage{lscape}
\usepackage{hyphenat}
\usepackage{adjustbox}

\usepackage{bookmark}
\usepackage{pifont}
\newcommand{\pmark}{\ding{67}}%
\usepackage{rotating}
\usepackage{forest}

\usepackage{tikz}
\usetikzlibrary{trees,positioning,shapes,shadows,arrows}

\begin{document}
\let\WriteBookmarks\relax
\def\floatpagepagefraction{1}
\def\textpagefraction{.001}
\shorttitle{Physical Fault Injection and Side-Channel Attacks on Mobile Devices: A Comprehensive Analysis}
\shortauthors{C. Shepherd, K. Markantonakis, N. van Heijningen, D. Aboulkassimi, C. Gaine, T. Heckmann, and D. Naccache}

\title [mode = title]{Physical Fault Injection and Side-Channel Attacks on Mobile Devices: A Comprehensive Analysis}

\author[1]{Carlton Shepherd}[orcid=0000-0002-7366-9034]
\cormark[2]
\ead{carlton.shepherd@rhul.ac.uk}
\address[1]{Information Security Group, Royal Holloway, University of London, Egham, Surrey, United Kingdom}
\address[2]{Netherlands Forensic Institute, The Netherlands}
\address[3]{Equipe Commune CEA Tech -- Mines Saint-Etienne, CEA Tech, Centre CMP, Gardanne, France}
\address[4]{French National Gendarmerie Research Center (CREOGN), France}
\address[5]{Institut de Recherche Criminelle de la Gendarmerie Nationale, France}
\address[6]{\'{E}cole Normale Sup\'{e}rieure, Paris, France}

\author[1]{Konstantinos Markantonakis}
\cormark[1]
\ead{k.markantonakis@rhul.ac.uk}

\author[2]{Nico {van Heijningen}}
\author[3]{Driss Aboulkassimi}
\author[3]{Cl\'{e}ment Gaine}
\author[4,5,6]{Thibaut Heckmann}
\author[4,6]{David Naccache}

\cortext[cor1]{Corresponding author}
\cortext[cor2]{Principal corresponding author}

\nonumnote{Part of this work was presented to the European Commission during the EU Horizon 2020 EXFILES project (No. 883156) ~\cite{exfiles_d51}.}

\begin{abstract}
Today's mobile devices contain densely packaged system-on-chips (SoCs) with multi-core, high-frequency CPUs and complex pipelines.  In parallel, sophisticated SoC-assisted security mechanisms have become commonplace for protecting device data, such as trusted execution environments, full-disk and file-based encryption. Both advancements have dramatically complicated the use of conventional physical attacks, requiring the development of specialised attacks. In this survey, we consolidate recent developments in physical fault injections and side-channel attacks on modern mobile devices. In total, we comprehensively survey over 50 fault injection and side-channel attack papers published between 2009--2021. We evaluate the prevailing methods, compare existing attacks using a common set of criteria, identify several challenges and shortcomings, and suggest future directions of research.
\end{abstract}



\maketitle

\section{Introduction}

Powerful personal mobile devices have become ubiquitous over the past decade, which is coupled with a substantial decline in the use of personal computers~\cite{statista:pc}. Estimates show that 3.8 billion people worldwide will be active smartphone users in 2021~\cite{statista:smartphones}. 84\% of US citizens use a smartphone regularly, rising to 96\% in the 18--29 year old category, while 52\% also use a tablet device~\cite{pew:mobile}. Mobile devices are used widely for security- and privacy-sensitive applications, including banking, instant messaging, navigation, social media, corporate email, and accessing cloud-based media. Consequently, device data is often of major interest to malicious adversaries and forensic investigators.

In certain scenarios, these actors may resort to physical attacks to circumvent existing software and hardware security mechanisms on mobile devices, such as trusted execution environments (TEEs), full-disk and file-based encryption, and secure boot procedures. In this work, we analyse two well-known classes of physical attacks---\emph{fault injections} and \emph{side-channel attacks}---and their application to mobile devices. Such attacks are well-understood in the smart card and secure element (SE) domain~\cite{markantonakis2009attacking,guilley2010fault,quisquater2001electromagnetic,kim2007faults}. Indeed, they are usually employed during the evaluation of smart cards and SEs under the Common Criteria framework~\cite{lomne2016common}.

Fault injection attacks (FIAs) perturb the device's physical conditions beyond that which it was intended; for example, using intense electromagnetic (EM) pulses, high ambient temperatures, and under- and over-volting the device's supply voltage. These attacks can induce errors in internal electronic components, which can be utilised to recover cryptographic keys and other secret data.  In contrast, physical side-channel attacks (SCAs) exploit physical measurements produced by the device during execution. Particular side-effects, like EM emissions and power consumption, may expose secret information if the side-effect is dependent on the data being executed. 

Unfortunately, the complexity of today's mobile devices has frustrated the application of traditional FIAs and SCAs used historically against simpler platforms, such as micro-controller units (MCUs).  Modern mobile system-on-chips (SoCs) employ high-frequency (1GHz+), multi-core CPUs with multi-stage pipelines and instruction-level parallelism; contain mixed analog and digital components; and are used to host complex operating systems (OSs) with various concurrent and virtualised processes. These aspects complicate the acquisition of side-channel measurements and the precise and accurate injection of faults. Ultimately, this has changed the types of physical attacks that have found success in the current literature.

In this work, we consolidate and extensively evaluate the state of the art in physical fault injection and side-channel attacks on mobile devices. In total, over 50 research papers are examined from 2009 to 2021, which are compared using the attack goals, prerequisites, target platforms, and reported results. From this work, we identify some significant challenges and limitations, and highlight potential future research directions. To the best of our knowledge, we present the first comprehensive survey in the area with respect to modern mobile devices.

\subsection{Scope}

This study examines \emph{physical} fault injection and side-channel attacks on SoC-based mobile devices that physically observe (SCAs) or perturb (FIAs) their execution, whether wholly or in part. In this context, mobile SoCs are defined as the primary integrated circuit (IC) for supporting a multitasking OS, e.g.\ Android, on a commercially available mobile platform. The SoC incorporates a single- or multi-core CPU with memory management unit (MMU) support; multiple memory modules, e.g.\ dynamic and non-volatile RAM, and flash storage; external peripheral interfaces, e.g.\ inter-integrated circuit (I2C); and, optionally, a dedicated GPU.

\begin{figure}
    \centering 
    \resizebox{\columnwidth}{!}{
    \begin{forest}
  for tree={rounded corners, top color=white, bottom color=gray!10, edge+={darkgray, line width=1pt}, draw=darkgray, align=center, anchor=children},
  before packing={where n children=3{calign child=2, calign=child edge}{}},
  before typesetting nodes={where content={}{coordinate}{}},
  [Physical Attacks,
    [Fault Injection\\Attacks (FIAs)
     [Invasive
        [Laser\\Beam]
        [
          [Focussed\\Ion Beam]
        ]
        [Heavy-Ion\\Micro-Beam]
      ]
     [Non-Invasive
        [Clock\\Glitches]
        [
            [Electromagnetic\\Fault Injections\\(EMFIs)]
            [Heating\\Attacks]
        ]
        [Voltage\\Glitches]
      ]
    ]
    [Side-Channel\\Attacks (SCAs)
      [Power\\Analysis]
      [
         [Acoustic\\Analysis]
         [Electromagnetic\\Analysis (EMA)]
      ]
     [Temperature\\Analysis]
    ]
  ]
\end{forest}
}
    \caption{Taxonomic scope of this survey.}
    \label{fig:taxonomy_scope}
\end{figure}
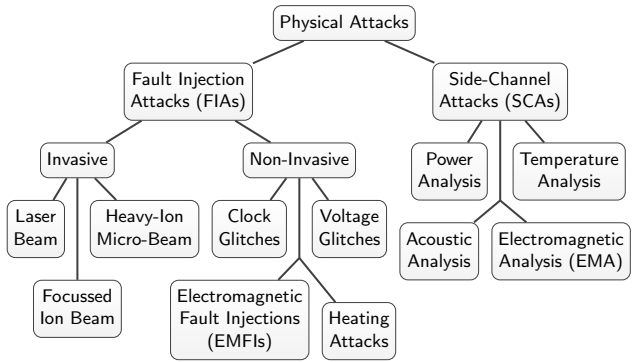

The taxonomic scope of this paper is shown in Figure \ref{fig:taxonomy_scope}. Wholly software attacks are outside the scope of this survey, which include but are not limited to: micro-architectural attacks exploitable only in software, such as cache timings and side effects of speculative execution, and software timing attacks on security protocols and cryptosystems. Along with alternative hardware attack vectors, such as NAND mirroring~\cite{skorobogatov2016bumpy} and hardware trojans~\cite{tehranipoor2010survey}, these topics constitute significant distinct research areas that justify separate analyses. 

While the focus of our work is physical FIAs and SCAs on mobile SoCs, references from the embedded systems literature are examined as supporting material where appropriate. This is particularly the case where a fault injection or side-channel attack has recently compromised a related evaluation target, like an Internet of Things (IoT) SoC, which may inspire future attacks on mobile devices.

\subsection{Related Surveys}

\begin{table*}
\centering
\caption{Summary of related surveys.}
\label{tab:fias}
\begin{tabular}{r|c|c}
\toprule
\textbf{Work} & \textbf{Year} & \textbf{Summary}  \\\midrule
Barenghi et al.~\cite{barenghi2012fault} & 2012 & Fault injection attacks on cryptographic systems.   \\
\rowcolor{gray!20}Shepherd et al.~\cite{shepherd2016secure} & 2016 & Security properties of secure execution platforms and TEEs for embedded systems.  \\
Fournaris et al.~\cite{fournaris2017exploiting} & 2017 & Micro-architectural attacks for embedded systems. \\
\rowcolor{gray!20}Spreitzer et al.~\cite{spreitzer2017systematic} & 2017 & Network, micro-architectural, software, and some hardware SCAs on mobile devices. \\
Mirzargar \& Stojilovi\'{c}~\cite{mirzargar2019physical} & 2019 &Physical SCAs and covert channels for FPGAs and cryptographic ASICs.\\
\rowcolor{gray!20}Sayakkara et al.~\cite{sayakkara2019survey} & 2019 & Electromagnetic side-channel analysis of embedded systems.\\
Han et al.~\cite{han2019side} & 2019 & Program analysis using side-channel analysis for embedded systems.\\\bottomrule
\end{tabular}
\label{tab:surveys}
\end{table*}

In prior work, Barenghi et al.~\cite{barenghi2012fault} (2012) presented a review of FIAs against symmetric and asymmetric cryptosystems on general computing systems. We note that this work precedes the proliferation of modern mobile SoCs and their associated security mechanisms, e.g.\ TEEs and full-disk encryption. In 2016, Shepherd et al.~\cite{shepherd2016secure} reviewed secure and trusted execution environments for IoT systems, comparing their security features and briefly discussing their susceptibility to physical attacks. No FIAs or SCAs were analysed in detail, however.  Spreitzer et al.~\cite{spreitzer2017systematic} (2017) surveyed general SCAs on mobile devices, covering physical attacks at a high level, alongside network traffic analysis, cache timings, and other software channels outside the scope of this work. Fournaris et al.~\cite{fournaris2017exploiting} (2017) focussed on X86 and ARM Rowhammer attacks, with a brief discussion of electromagnetic analysis (EMA). In 2019, Mirzargar and Stojilovi\'{c}~\cite{mirzargar2019physical} surveyed SCAs on field-programmable gate arrays (FPGAs) and application-specific ICs (ASICs). The use of power, electromagnetic, and thermal side-channels were examined for creating covert channels.  Sayakkara et al.~\cite{sayakkara2019survey} (2019) presented a tutorial on using EMA for digital forensics. Methods are described for EMA measurement acquisition, their analysis for hardware and software profiling, and the related standards; it is unrelated to attacks on mobile security mechanisms. Han et al.~\cite{han2019side} (2019) presented a tutorial of program analysis using EMA and power analysis. The work provided a description of methods for signal acquisition and their statistical analysis using correlation analysis and machine learning.

These works are summarised in Table~\ref{tab:surveys}. Additionally, other studies have surveyed general mobile attacks~\cite{vidas2011all}, cryptographic analysis methods~\cite{breier2014survey}, and attacks on simpler security devices, e.g.\ smart cards~\cite{markantonakis2009attacking}. However, no survey paper has extensively covered state-of-the-art physical fault injection and side-channel attacks on modern mobile devices. This paper addresses that gap.

\subsection{Structure}

We begin in \S\ref{sec:prelim} with background information on mobile SoCs, their high-level system features, and their use in common hardware-assisted security mechanisms. Following this, comprehensive surveys of physical fault injection and side-channel attacks are presented in \S\ref{chapter:fi} and \S\ref{chapter:sca} respectively. The surveyed works are extensively compared in \S\ref{sec:eval}, where several shortcomings, challenges, and potential future research directions arising from existing work are identified. Finally, \S\ref{sec:conclusion} concludes this paper.

\section{Background}
\label{sec:prelim}

This section provides background information for understanding recent fault injection and side-channel attacks. In summary, we provide an overview of mobile SoCs and widely deployed hardware-assisted security mechanisms, including TEEs, secure boot, key management systems, and full-disk and file-based encryption.

\subsection{Mobile System-on-Chips (SoCs)}

System-on-chips---ICs with the core components for establishing a working computing system---have become the centrepiece of contemporary mobile hardware due to their reduced energy efficiency and physical footprint. Silicon vendors typically design SoCs through the configuration of reusable semiconductor intellectual property (IP) blocks developed in-house or licensed from a third party, like ARM.  Vendors integrate varying numbers of such components according to their specifications, which are fabricated as interconnected subsystems onto a single IC using specialised buses for data transfer and control.  High-frequency multi-core CPUs with multi-level cache hierarchies and MMUs are at the foundation of today's application processors, e.g.\ ARM's Cortex-A73~\cite{arm:cortex73}. Supporting on-SoC components include diverse memory units, e.g.\ dynamic RAM, flash storage and read-only memory; timers and real-time clocks; security extensions and cryptographic accelerators, e.g.\ ARM TrustZone and CryptoCell~\cite{arm:layered_security}; integrated GPUs and audio digital signal processing (DSP) units; and input/output (I/O) interfaces for communicating with sensor hubs, input devices, and other off-SoC peripherals.

The exact capabilities of mobile SoCs varies between silicon vendors. At the time of publication, ARM-based SoC designs are used in approximately 90\% of mobile handsets, tablets, and IoT devices~\cite{arm:statista}.  The Qualcomm Snapdragon, HiSilicon Kirin, Samsung Exynos, Apple A, and the MediaTek Helios series are some widely used ARM-based SoCs used by mobile original equipment manufacturers (OEMs).

\subsection{Trusted Execution Environments (TEEs)}
\label{sec:mobile_tee}

A TEE is an isolated, parallel execution environment that aims to protect sensitive code and data from privileged software attacks. TEEs have gained widespread use on mobile devices for preserving user authentication algorithms, cryptographic keys, digital rights management systems, and payment processing applications.  At present, ARM TrustZone and the GlobalPlatform TEE---a suite of specifications governing the use of TEEs---are the main enablers on mobile devices, arising from their compatibility with the ARM architecture~\cite{arfaoui:tee,ekberg2014untapped,shepherd2016secure}. Intel Software Guard Extensions (SGX) and the AMD Platform Security Processor (PSP) are two alternative TEE systems for Intel and AMD chipsets; however, these are independent technologies on predominantly non-mobile platforms, e.g.\ servers and workstations.

\subsubsection{GlobalPlatform TEE Specifications}

\begin{figure}
\centering
\includegraphics[scale=1,width=\linewidth]{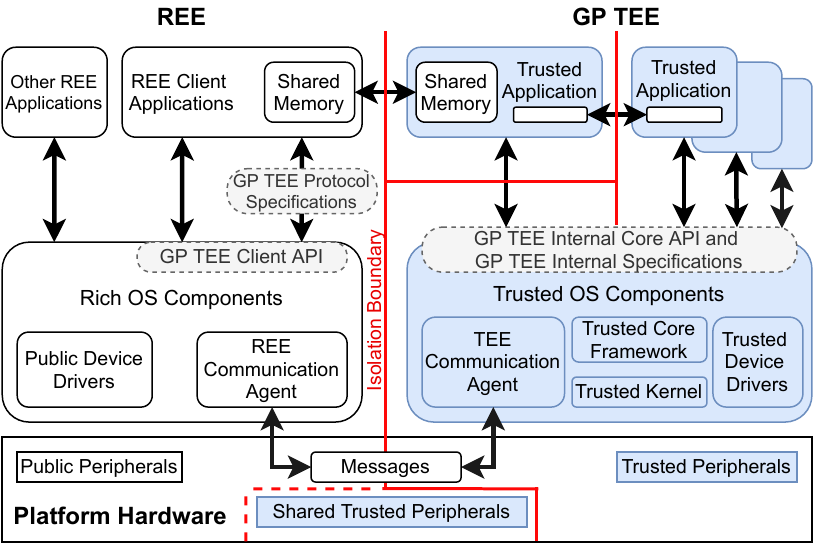}
\caption{GlobalPlatform TEE system architecture~\cite{gp:arch}.}
\label{fig:gp_tee}
\end{figure}

The GlobalPlatform TEE (GP TEE) is a suite of specifications for defining the architecture, management and security requirements of a TEE~\cite{gp:arch,gp:tmf,gp:pp}.  It divides execution into two worlds: the rich execution environment (REE, or the \emph{untrusted} or \emph{non-secure} world) and the TEE (\emph{trusted} or \emph{secure} world), as shown in Figure~\ref{fig:gp_tee}. The REE includes the mobile's native OS, its firmware and its user applications, whereas the TEE contains a separate trusted OS with a security kernel, device drivers for security-critical peripherals, and a set of lower-privileged trusted applications (TAs). Each world must execute with hardware-enforced isolation if they utilise the same underlying hardware platform. This to impede privileged REE software adversaries from compromising the control flow, memory regions and peripheral devices of the secure world.

For inter-world communication, REE applications may access the TEE using a tightly controlled interface at the platform's highest privilege level or, alternatively, using a shared memory view created by the TEE. The GlobalPlatform Client API~\cite{gp:client} standardises the communication interfaces through which REE applications can invoke TEE TA functions and transmit/receive data, including supported data types and response codes. The Internal Core API~\cite{gp:internal} specifies functions through which TAs can access services provided by the trusted OS, such as sources of time, external peripherals, cryptographic functions, and secure storage media. Additionally, the Secure Element~\cite{gp:ses}, Sockets~\cite{gp:sockets} and Trusted User Interface~\cite{gp:tui} APIs have also been specified for enabling TAs to access hardware secure elements, networking services, and user interface devices respectively. For the full specifications, the reader is referred to the GlobalPlatform Technology Document Library~\cite{gp:tech_lib}.

The GP TEE Protection Profile (PP) characterises the security definitions and requirements for evaluating a target of evaluation (TOE) under the Common Criteria (CC) framework. It addresses \emph{``threats to the TEE assets that arise during the end-usage phase and can be achieved by software means...focuses on non destructive software attacks that can be easily widespread...and constitute a privileged vector for getting undue access to TEE assets without damaging the device itself''}~\cite{gp:pp}.  At a minimum, it stipulates that TEEs should defend against two high-level adversary types:
\begin{itemize}
    \item \textbf{Basic remote attacker}:\ \emph{``Performs the attack on a remotely controlled device or alternatively makes a dow-nloadable tool that is very convenient to end-users. The attacker retrieves details of the vulnerability identified in the identification phase and} [...]\ \emph{makes a remote tool or malware and uses techniques such as phis-hing to have it downloaded and executed by a victim} [into the untrusted world]\emph{.''}
    \item \textbf{Basic on-device attacker}:\ \emph{``Has physical access to the target device; it is the end-user or someone on his behalf.  The attacker is able to retrieve exploit code, guidelines written on the internet on how to perform the attack, and downloads and uses tools to jailbreak, root, reflash the device in order to get privileged access to the REE allowing the execution of the exploit. The attacker may be a layman or have some level of expertise but the attacks do not require any specific equipment.''}~\cite{gp:pp}.
\end{itemize}

The TEE OS and TAs are deemed trusted software components; security issues within either can compromise the services that the TEE aims to provide.  From a hardware perspective, the GP TEE should have access to a secure clock, entropy source, and volatile and non-volatile memory.  The GP TEE must also be initialised from a root of trust (RoT) using a secure boot process to ensure authenticity. How this is implemented in practice is both platform- and architecture-specific. We describe this process for ARM TrustZone in \S\ref{sec:secure_boot}, the main method by which GP TEEs are instantiated on ARM-based mobile SoCs. Crucially, however, the TEE's initialisation process must be performed without trusting or co-operating with the REE.  In general, the GP TEE must operate self-sufficiently without depending on any untrusted components in the REE~\cite{gp:arch,gp:pp}.

GlobalPlatform-compliant TEEs must satisfy Common Criteria Evaluation Assurance Level 2 (EAL2) as a minimum baseline. In general, TOEs that meet this level provide `low-to-moderate' security assurances under the CC grading framework and is satisfied through evidence of structured testing~\cite{cc:definitions}. This is lower than smart cards and secure elements, which are usually evaluated to CC EAL4+ thereby providing `moderate-to-high' security assurances. EAL4+ demands a higher degree of security testing rigour, requiring methodical design, testing and review in addition to the requirements of lower EALs. 

It is important to consider that smart cards and secure elements often consider physical attackers with access to specialised testing equipment, such as high-end oscilloscopes, chemical workbenches, and electromagnetic pulse apparatus~\cite{tunstall2017smart}. These threats are beyond the minimum protection scope of GlobalPlatform-compliant TEEs.\footnote{It is still possible that GlobalPlatform TEE implementations may contain countermeasures against advanced physical attacks beyond the minimum compliance requirements.} This lack of coverage means that physical fault injection and side-channel attacks can offer a potential avenue for subverting GP TEE implementations.

\subsubsection{ARM TrustZone}
\label{sec:trustzone}

\begin{figure}[h]
\centering
\includegraphics[scale=1,width=\linewidth]{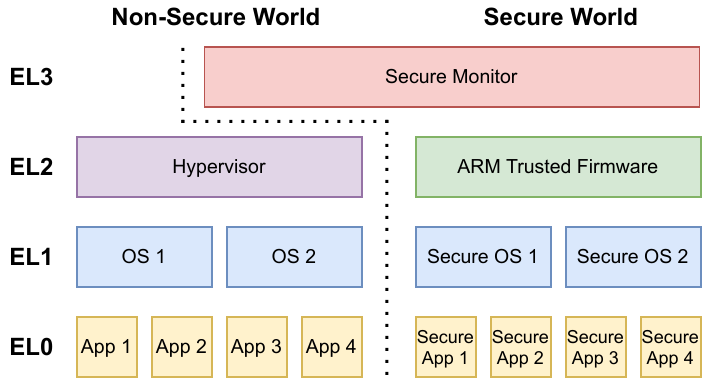}
\caption{The ARM v8.4-A exception model~\cite{arm:arm84}.}
\label{tz:exception_levels8.4}
\end{figure}

ARM TrustZone is a set of extensions to the ARM architecture for creating a secure world of execution. It is the main implementation for providing the hardware-enforced separation required by GlobalPlatform-compliant TEEs on ARM-based devices~\cite{arfaoui:tee,ekberg2014untapped,shepherd2016secure}. Similarly, TrustZone partitions the execution platform into `secure' and `non-secure' worlds.  Each world hosts its own OS and applications at different processor exception levels, which communicate over a secure monitor at the highest ARM exception level (EL3), as shown in Figure \ref{tz:exception_levels8.4}. Secure monitor mode is entered into using the ARM Secure Monitor Call (SMC) interface, which handles the world context switch. A reference implementation of the secure monitor code and bootloading process is provided by the Trusted Firmware project~\cite{trusted_framework}, which manufacturers may tailor to their own SoCs.

A control signal, the `non-secure' (NS) bit, is used by TrustZone-enabled SoCs to store the current world of execution. The NS-bit is propagated through all areas of the SoC where adversaries may attempt to access secure world material, including page tables, cache lines, and bus transactions to memory firewalls and peripheral controllers. Unauthorised access attempts to secure world-only resources are prevented by on-SoC controllers that are configured at boot-time. As specific examples, the TrustZone Peripheral Controller (TZPC) prevents non-secure accesses to secure world-only peripheral interfaces, and the TrustZone Address Space Controller (TZASC) can be used to prevent non-secure accesses to secure world memory addresses. 

At its core, TrustZone is a SoC security architecture; it does not provide the software components for realising a complete TEE containing a hardened security kernel, support for multiple TAs, and so on. To bridge this gap, numerous TrustZone-based TEE implementations have been developed commercially, e.g.\ Huawei's iTrustee~\cite{itrustee}, Samsung's TEEGRIS~\cite{teegris}, Trustonic's Kinibi~\cite{trustonic}, and the Qualcomm TEE~\cite{qualcomm:qtee}, many of which conform to the GlobalPlatform TEE specifications. 
Lastly, we note that TrustZone aims primarily to defend against privileged non-secure world software attacks. Hardware threats are outside of its protection scope; if such attacks are considered reasonable, ARM's SecurCore smart card may be used with a TrustZone-enabled SoC to protect the relevant assets~\cite{tz:threatmodel}.

\subsubsection{Apple Secure Enclave}
\label{sec:proprietary}

The Secure Enclave is an independent subsystem on Apple SoCs (from the Apple A7~\cite{apple:ios_security}) for protecting content even when the application processor is compromised. Detailed technical data is not publicly available besides high-level features described in the Apple security documentation~\cite{apple:ios_security,blackhat:ios,apple:support_sep} and outcomes from independent reverse engineering efforts~\cite{mandt2016demystifying,duo:sep}. From the available information, the Secure Enclave comprises a hardware co-processor---the Secure Enclave Processor (SEP)---a true random number generator (TRNG); a unique root key for implmenting device-specific key binding; an I2C controller for secure storage to external non-volatile memory (NVRAM); and a public key accelerator (PKA). The Secure Enclave manages cryptographic operations for Apple iOS, including file-based encryption, and user data for the Touch ID and Face ID biometric systems~\cite{apple:ios_security}. 

The SEP is a physically separated processor that runs a security kernel---sepOS, an Apple fork of the L4 micro-kernel---which is initialised during the secure boot process. This separation offers inherent resistance to certain classes of side-channel analysis from software adversaries on the application processor, e.g.\ cache timing attacks. The confidentiality of run-time Secure Enclave data is protected using a Memory Protection Engine, which encrypts memory blocks using AES in XEX (xor-encrypt-xor) mode and computes cipher-based message authentication code (CMAC) tags that are both stored in external untrusted DRAM. 

The PKA is used to secure public-key cryptographic operations, namely key generation (i.e.\ from the root SEP key), encryption and decryption, and digital signature services using RSA and elliptic curve cryptography (ECC). Undisclosed countermeasures are implemented to resist power analysis attacks, including differential power analysis (DPA).  Besides this, a separate AES memory encryption engine, located in the DMA path between the application processor and NAND flash memory, is used for file-based encryption---discussed in \S\ref{sec:fde}---the keys for which are managed by the Secure Enclave. From the Apple A9 SoC series, undisclosed countermeasures are also utilised by this module to thwart power analysis attacks. Related to this, the Secure Enclave contains monitoring circuits to enforce the intended operating voltage and clock frequency against fault injection attacks~\cite{apple:security_guide}.

\subsubsection{Google Titan M}
\label{sec:google_titan}

In 2017, Google announced the Titan M for the Pixel 2 smartphone, a mobile hardware security module that executes independently of the application SoC~\cite{google:titan}. Similar to the Apple Secure Enclave, it is also largely undocumented. From what is known, the Titan contains flash memory, CPU, RAM and a TRNG in a single package designed to resist physical side-channel analysis, including power and electromagnetic analysis, and voltage-, clock- and temperature-based fault attacks.  The Titan M supports general key management and cryptographic operations using the StrongBox KeyStore APIs (\S\ref{sec:gm_ta}) and is used for flash memory encryption/decryption. In addition, the Titan stores the last known safe Android version during legitimate device updates. This is referenced by the Pixel's secure boot process for preventing rollback attacks to vulnerable OS versions.

\subsection{Secure Boot}
\label{sec:secure_boot}

\begin{figure}
\centering
\includegraphics[scale=1,width=\linewidth]{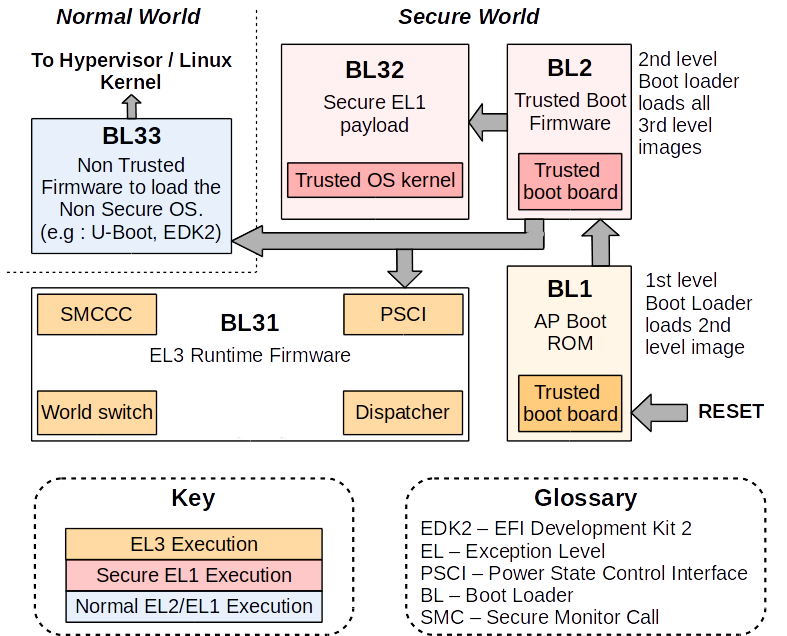}
\caption{Generic ARM boot procedure~\cite{arm:soc,virtualopensystems}.}
\label{tz:tfbf}
\end{figure}

Adversaries with the ability to load unauthorised bootloaders and secure/non-secure world binaries may trivially disable critical security mechanisms.  Consequently, secure boot procedures have  been deployed for cryptographically enforcing the use of authorised software components at boot-time. Mobile devices contain multiple bootloader stages of increasing complexity and decreasing security privilege; a generic boot procedure is shown in Figure~\ref{tz:tfbf}.  At the beginning, code within on-SoC ROM (BL1) is used to perform basic SoC setup operations, e.g.\ power-on self-test (POST) and clock initialisation, after receiving a reset signal.  The trusted world binary and secure monitor firmware (BL2) are subsequently authenticated and loaded thereafter. Importantly, this is done \emph{before} loading the non-secure world image containing the native OS and its applications (BL3). 

\begin{figure}
\centering
\includegraphics[scale=1,width=\linewidth]{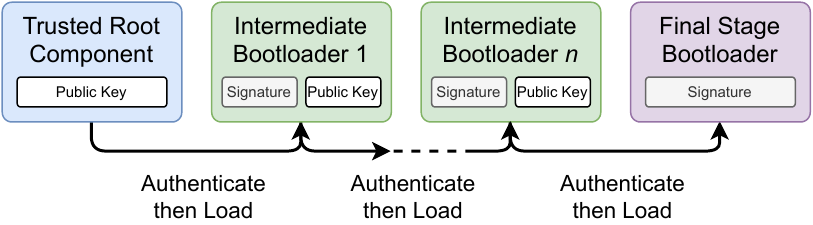}
\caption{High-level authenticated secure boot process.}
\label{fig:auth_boot}
\end{figure}

For authenticating each component, a common method involves verifying a sequence of digital signatures from the initialisation ROM, which is inherently trusted. The ROM contains a public key certificate---a hash of which may reside in one-time programmable (OTP) eFuses---for verifying the signature of the second bootloader.  Upon successful verification, this bootloader is loaded, which itself contains a public key for verifying and loading the following bootloader. The process is repeated until the non-secure world image is eventually loaded and executed (Figure~\ref{fig:auth_boot}).


Boot images are signed under a private OEM signing key; if any unauthorised components are introduced, the signature verification check will fail and the boot sequence terminated. Some devices also contain OTP eFuses that are permanently set (`blown') after failed checks, thus signalling that device assets may have been compromised~\cite{knox_warranty}. Details are scarce about the exact cryptographic algorithms used by OEMs, although the ARM Trusted Board Boot requirements specify SHA-256 and RSA-2048 or 256-bit ECDSA for bootloader signatures~\cite{arm:soc}.

\subsection{Secure Storage and Key Management}
\label{sec:gm_ta}

TEEs may utilise one of two methods for secure persistent data storage: 1), encrypting data using keys accessible only to the TEE; and, 2), directly storing data to an external device over a trusted path. To support the first case, TEE implementations usually possess their own key hierarchies rooted in a device-specific hardware unique key (HUK) from which descendent keys are generated using a key derivation function (KDF). These keys may then be used by TAs and the TEE OS to encrypt arbitrary binary data, including other cryptographic keys, whether generated within the TEE or sent from the REE. This provides a notion of device `binding' where data cannot be transferred and decrypted on another device with its own HUK. GP TEEs support a variety of standard cryptographic algorithms, such as 128/192/256-bit AES in various modes of operation, e.g.\ CBC, XTS, and CTR; up to 2048-bit RSA; 160- to 521-bit ECC (ECDSA and ECDH); and 128- or 196-bit triple DES. The Internal Core API~\cite{gp:internal} specifies a full list of supported algorithms and modes of operation.  

At present, regular application developers are not given direct access to TEEs. OEMs have, however, developed key management systems for providing abstracted access to TEE-based cryptography services. On Android devices, the Keymaster service exposes API functions for generating, using and containerising cryptographic keys in the TEE from user applications~\cite{android:keystore}. Key material is generated within the TEE and is never exposed to the calling application. Developers are purposefully not given access to these keys; instead, they indirectly use this material by calling abstracted encryption, decryption, and signature verification and signing functions available through the Android framework.  Where supported, the StrongBox Keymaster serves as a TEE substitute for addressing physical attacks, which implements the Keymaster in a mobile hardware security module, e.g.\ the Titan M~\cite{google:strongbox} (\S\ref{sec:google_titan}).

On Apple devices, the SEP (\S\ref{sec:proprietary}) is manufactured with a hardware-based unique ID (UID) from which cryptographic keys are derived for device binding. These keys are never exposed to the main application processor.  The SEP handles key management for secure boot signature verification, securely storing biometric data at rest, and general OS cryptographic operations. For the latter, additional keys are derived from a user-inputted password and a hardware secret, which are used for per-file and volume encryption~\cite{apple:ios_security}.

\subsection{Full-Disk and File-Based Encryption}
\label{sec:fde}

Full-disk encryption (FDE) encrypts all user data on a device at a block level using symmetric key encryption, usually AES. Throughout standard device usage, data is continuously encrypted before it is written to a flash storage device that exposes itself as blocked memory; during read operations, the data is decrypted before returning it to the parent process.  Android's FDE implementation is based on the \texttt{dm-crypt} Linux kernel module, which encrypts data using a device disk encryption key (DEK) with 128-bit AES in CBC mode. The DEK is encrypted by a key encryption key (KEK) derived from the user's PIN or password, both of which are managed within the TEE. Only if the user successfully passes an authentication challenge are the keys released; failing this, neither the KEK nor the DEK are unlocked, thereby preventing data decryption~\cite{android:fde}.  

File-based encryption (FBE) encrypts data at a filesystem level, unlike FDE that operates at a block or volume level. FBE is generally considered to be FDE's successor on mobile devices: it permits essential data to still be used for critical device functions---accessibility services, emergency calls, and alarm managers---while keeping sensitive data encrypted when the device is locked~\cite{android:fbe}. Commonly, FBE generates unique AES file encryption keys (FEKs) for dynamically encrypting and decrypting user files, e.g.\ documents, as they are written to and read from a persistent storage device. FBE is used by Secure Enclave-enabled Apple devices, which manages keys within the Apple SEP. In contrast, Android devices (version 7.0+) support directory-level encryption using keys managed in the Keymaster TEE TA or a mobile hardware security module. 
\section{Fault Injection Attacks}
\label{chapter:fi}

Fault injection attacks (FIAs) are active attacks that physically perturb the device beyond its intended operating conditions. This can induce abnormal system behaviour for uncovering secret data under execution or accessing restricted code regions and functionality.   
FIAs have a long-standing history and have been analysed in great depth in the wider literature, particularly for smart cards and other embedded MCUs~\cite{barenghi2012fault,anderson1997low,boneh1997importance,yen2000checking,bao1997breaking,page2006fault,hemme2004differential}. 

FIAs are categorised as \emph{transient vs.\ permanent}, and \emph{invasive vs.\ non-invasive}.  Transient faults are temporary errors that are recoverable following a system reset or cessation of the fault source. Their aim is to (temporarily) disrupt the program control flow or corrupt the results of an instruction to gain unauthorised access to sensitive code and data. While having the same end goal, permanent faults indefinitely alter the state of target components, the effects of which persist irrespective of device restarts and resets. 

\begin{figure}
    \centering
    \includegraphics[width=0.88\linewidth,interpolate=on]{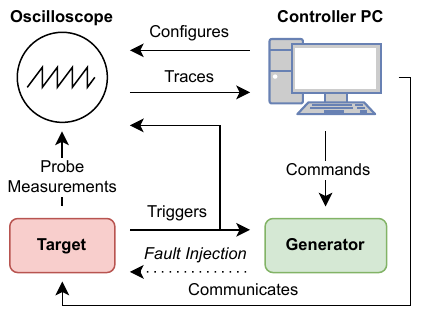}
    \caption{Generic FIA and SCA setup. For FIAs, the generator injects the fault after receiving the trigger; for SCAs, the trigger activates the measurement process.}
    \label{fig:attack_setup}
\end{figure}

A typical physical FI (and SCA) setup is shown in Figure \ref{fig:attack_setup}. The investigator instruments the device under test (DUT) to activate an external generator at run-time following a hardware- or software-based trigger. The generator perturbs the DUT with a precisely timed and calibrated fault using the desired method; for example, an EM burst over a particular SoC die location or a momentarily high voltage at the DUT's power supply.\footnote{Instrumenting device triggers is a significant challenge in black-box environments, rendering accurate and precisely calibrated fault injections more difficult than white-box testing.} The investigator uses a control computer to analyse the resulting outputs---register values, memory address contents, encryption/decryption results, oscilloscope traces, etc.---in order to discover useful faults, such as instruction corruption and decryption errors. 

\subsection{Invasive FIAs}

Invasive FIAs involve significant alteration to the DUT, whether during the attack's preparation or execution. Relevant preparation techniques include depackaging the SoC or IC and removing any protective layers (decapsulation) to directly induce faults into its internal components. These processes risk irreparable damage or destruction of the target under evaluation, potentially leading to permanent data loss. At the execution stage, the application of high-energy visible light and ultraviolet pulses, near-infrared lasers, and focussed ion beams may be used to flip the transistor states of memory cells and other components~\cite{skorobogatov2002optical,van2011practical,pouget2008dynamic,schmidt2009optical,torrance2009state,li2015heavy,vasselle}.  Such methods have been historically successful against simpler systems~\cite{kim2007faults,barenghi2012fault,van2011practical}, but have found limited utility against mobile SoCs from publicly available information.

\subsubsection{Ion-based FIAs}

The costliest FIAs employ focussed ion beams (FIB)---typically liquid metal ion sources, e.g.\ gallium ions---and heavy-ion microbeams (HIMs) that operate with extreme precision ($\sim$2.5nm). In the semiconductor industry, FIB workstations permit the manipulation of IC silicon substrates \emph{in situ}, potentially enabling hardware-level reworking (microsurgery) of security-critical components and the reading of hardware-fused keys. In 2009, Torrance and James~\cite{torrance2009state} used FIBs to perform IC microsurgery for reading encryption keys from an application-specific IC (ASIC) implementation of the DES cipher. In comparison, HIMs ionise IC semiconductor material with high-energy radiation to induce single-event upsets (SEUs) in digital logic systems. This was leveraged by Li et al.~\cite{li2015heavy} (2015) to trigger exploitable decryption faults in an RSA implementation on an SRAM FPGA. The FPGA was irradiated using 1 GeV of carbon ions that generated decryption bit errors at a rate of 8 times per hour. The authors theorise that 1024-bit private key recovery can be achieved after $\sim$128 hours, but experimental results were not given. FIB and HIM attacks impose substantial cost (\$100,000+ USD) and require access to specialised expertise and testing equipment. No such attacks have been publicly disclosed against mobile SoCs to date. 

\subsubsection{Laser FIAs}

Laser FIAs target ICs with intense sources of visible, infrared or ultraviolet light with the aim of inducing SEUs. In 2017, Vasselle et al.~\cite{vasselle} presented the results of a laser FIA that bypassed the secure boot process of an undisclosed Android smartphone with an ARM Cortex A9-based (1.4GHz) SoC. This was achieved following physical and software reverse engineering, package decapsulation, and targeting the SoC with a 978nm infrared laser. The authors could modify the Secure Configuration Register (SCR) containing the TrustZone NS-bit with 95\% accuracy (see \S\ref{sec:trustzone}) in order to access secure world assets during non-secure execution.

Beyond mobile SoCs, Colombier et al.~\cite{colombier2019laser} (2019) described laser FIAs for modifying the bits of words fetched from flash memory on a ChipWhisperer board with a 32-bit ARM Cortex-M3 (7.4MHz). The MCU was decapsulated before applying a 3W infrared laser (1,064 nm), which generated faults that modified \texttt{movw} ARM instructions to \texttt{movt}.\footnote{The \texttt{movw} instruction moves the least significant 16 bits of an immediate integer value into a target register's lowest 16 bits, while \texttt{movt} moves the \emph{most} significant bits to a register's \emph{upper} 16 bits.} Using these errors, 128-bit AES key recovery was possible from a reference AES software implementation after 128 attempts and two faulty ciphertexts. Additionally, the faults were also used to bypass the authentication check of a mock PIN verification algorithm.

\subsection{Non-Invasive FIAs}

Non-invasive FIAs require little-to-no tampering of the DUT, e.g.\ SoC/IC decapsulation, removing adjacent components, and the use of corrosive chemicals. With reasonable care, their effects usually disappear after removing the stimulus or resetting the device. As a rule of thumb, such attacks can be conducted using lower cost, commercially available equipment and risk significantly less damage than their invasive counterparts~\cite{barenghi2012fault}. Four well-studied classes of fault injection attacks fall into this category:
\begin{itemize}
    \item \textbf{Voltage-based glitch attacks}: Under- or over-volt the device's power supply beyond its intended operating limits~\cite{qiu2019voltjockey,barenghi2010low,timmers2017escalating,controlling_pc_fi,bypassing_secure_boot,barenghi2009low,murdock2020plundervolt} (\S\ref{sec:v_fi}).
    \item \textbf{Clock glitches}: Increase or decrease the device's clock frequency used to synchronise the operation of internal components~\cite{blomer2014practical,agoyan2010clocks,tang2017clkscrew,korak2014effects} (\S\ref{sec:c_fi}).
    \item \textbf{Heating attacks}: Expose the DUT to temperatures beyond its maximum or minimum guidelines~\cite{korak,kumar2014precise,govindavajhala2003using} (\S\ref{sec:heating}).
    \item \textbf{Electromagnetic FIs} (EMFIs): Expose DUT components to targeted, high-energy EM pulses~\cite{dehbaoui2013electromagnetic,quisquater2002eddy,elmohr} (\S\ref{sec:emfi}).
 \end{itemize}

Table~\ref{tab:fia_comp} presents a summary of FIAs and their temporal and spacial precision, relative complexity, and damage risk. 
\begin{table*}
\label{tab:fia_comp}
\centering
\caption{Summary of FIAs based on \cite{barenghi2012fault} and \cite{li2015heavy}.}
\begin{threeparttable}
\begin{tabular}{r|c|c|c|c|c}
\toprule
\textbf{Technique} & \textbf{Precision (Space)} & \textbf{Precision (Time)} & \textbf{Cost} & \textbf{Skill} & \textbf{Damage Risk} \\\midrule
 
 Voltage Glitch & Low & Moderate & Low & Moderate & Low \\
\rowcolor{gray!20} EMFI & Moderate & Moderate & Moderate &Moderate& Low \\
 Clock Glitch & Low & High & Low & Moderate & Moderate \\
\rowcolor{gray!20} Heating & Low & Low & Low & Low & Moderate \\
 Light Pulse & Moderate & Moderate & Moderate &High& Moderate \\
\rowcolor{gray!20} Laser Beam & High & High & High & High & High \\
 FIB & Very High & Very High & Very High & Very High &High \\
\rowcolor{gray!20} HIM & Very High & Very High & Highest & High &High \\\bottomrule
\end{tabular}
\begin{tablenotes}
\item \textbf{EMFI}: Electromagnetic FI, \textbf{FIB}: Focussed ion beam, \textbf{HIM}: Heavy-ion micro-beam.
\end{tablenotes}
\end{threeparttable}
\end{table*}

\subsubsection{Voltage-based Glitch Attacks}
\label{sec:v_fi}
Voltage glitches manipulate the target's supply voltage. An attacker can generate single- or multi-bit faults by under- or over-volting this source or redirecting it to ground to generate brownouts. This can corrupt the contents of memory units or coerce microprocessors into misinterpreting and even skipping program instructions.  In existing literature, such effects have been used to skip security-critical checks, e.g.\ digital signature verification; bypass system-level access control features; and to recover information about cryptographic key material under execution.

One of the first voltage glitches against a mobile phone SoC was presented by Barenghi et al.~\cite{barenghi2009low} (2009), who targeted a reference software RSA implementation on a 32-bit ARM9 CPU (ARM926EJ-S).\footnote{We note that the target is a single-core, low-frequency (266 MHz) SoC, which can be considered obsolete nowadays.} The work exploited errors in energy-intensive load instructions after the target was provided a low supply voltage (underfeeding). During instruction fetches, the authors could change the binary encoding of logical operations (\texttt{AND} to \texttt{EOR}), conditional additions (\texttt{ADDNE} to \texttt{ADDEQ}), and conditional branches (\texttt{BNE} to \texttt{BEQ}). These errors were used in three attacks against OpenSSL v0.9.1i: 1), for RSA factorisation when using the Chinese remainder theorem (CRT); 2), an e-th root attack for recovering an input message encrypted under a correct and faulted encryption procedure; and 3), a theoretical attack for secret key recovery during message signing. For the first, 6.6\%--39\% RSA-CRT computations could be faulted of which 3.0\%--39\% were exploitable, while, for the second, 36.42\%--62.77\% of 1,000 faulted instances were useful. The instruction swaps occurred only a small number of times, which \emph{``may be reduced up to a single one in the whole computation of a target algorithm,''} thus rendering the third attack impractical.  

Later, Barenghi et al.~\cite{barenghi2010low} (2010) used the same method to recover AES round sub-keys using differential fault analysis (DFA). The approach is independent of the key length and the number of rounds, and requires both fault-free and faulted ciphertexts from the same plaintext. The authors interrogated the same SoC from \cite{barenghi2009low}, which executed Linux 2.6 and a software AES implementation based on OpenSSL.  After experimental analysis, key recovery could be achieved following 100,000 encryptions with different plaintexts and 2,000,000 encryptions of the same plaintext.

\begin{figure}
    \centering
    \includegraphics[width=0.72\linewidth]{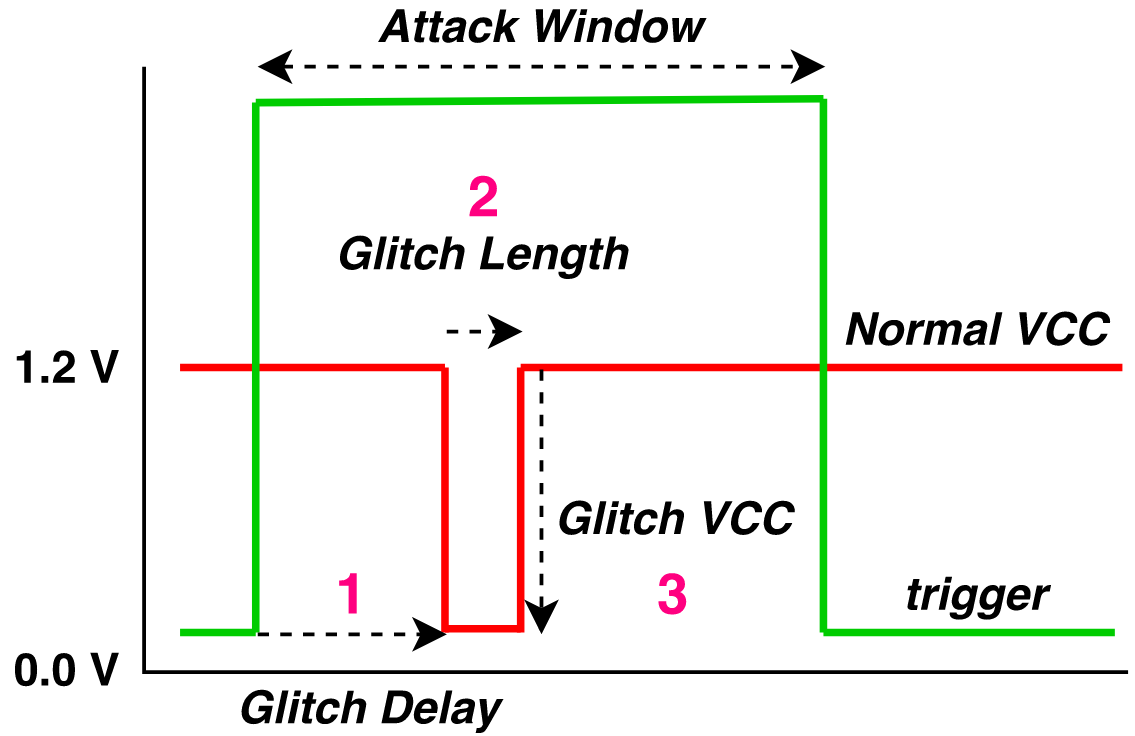}
    \caption{Voltage FIA parameters~\cite{timmers2017escalating}.}
    \label{fig:glitch_params}
\end{figure}

In 2016, Timmers et al.~\cite{controlling_pc_fi} demonstrated two voltage FIAs for privilege escalation on a Xilinx Zynq-7010 SoC (ARM Cortex-A9). Both attacks exploited instruction corruption vulnerabilities with two load instructions (\texttt{LDR} and \texttt{LDMIA}) on the 32-bit ARM architecture (AArch32), which occurred when under-volting the device. The first FIA targeted the BL1 bootloader (see \S\ref{sec:secure_boot}) to gain code execution privileges within the TEE. It required the BL1 image to be initially overwritten with a malicious payload containing shellcode and relevant pointers; the fault was then precisely injected after shellcode was copied into volatile memory and during the copying of the pointers. On successful occasions, the instruction corruption caused the processor to copy the pointer into its program counter (PC) register, therefore executing the shellcode and transferring control flow to the attacker. The second method used the same approach as an exploit delivery mechanism in situations where the REE and TEE communicate over a shared memory buffer. The fault was injected after loading the shellcode during the world context switch, causing the CPU to load the shellcode pointer into the PC register thereby triggering its execution in the TEE. In total, 10,000 fault attacks on the \texttt{LDR} and \texttt{LDMIA} instructions were conducted. Only a single (0.01\%) corruption occurred using \texttt{LDR}, whereas 27 glitches were successful (0.27\%) against \texttt{LDMIA}. Neither attack was mounted against TEEs or secure boot processes on an OEM mobile device.

\begin{figure*}
    \centering
    \includegraphics[width=0.66\linewidth,interpolate=on]{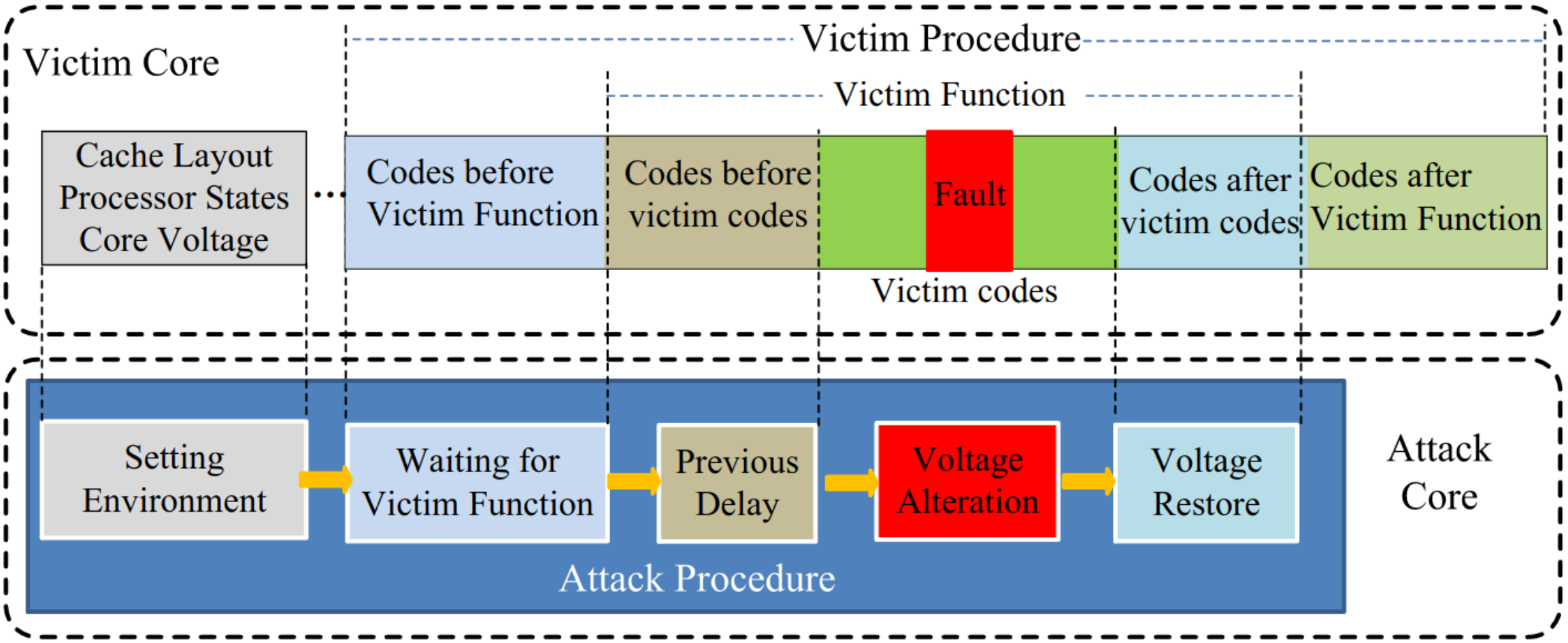}
    \caption{\centering VoltJockey attack sequence~\cite{qiu2019voltjockey}.}
    \label{fig:voltjockey}
\end{figure*}

In 2017, Timmers and Mune~\cite{timmers2017escalating} demonstrated voltage FIAs for Linux-based privilege escalation on an undisclosed ARM Cortex-A9-based SoC. After determining the glitch parameters---the voltage, length, and delay (Figure \ref{fig:glitch_params})---the authors targeted the \texttt{open} syscall when an unprivileged application attempted to access physical memory via \texttt{/dev/mem/}. The application was instrumented to trigger the fault during the kernel's access control check, which caused it to be skipped. The second FIA used the instruction corruption attack from \cite{controlling_pc_fi} to change the CPU PC register to a predetermined address while executing random kernel syscalls, generating system crashes as a proof of concept. The third attack used an FIA to bypass the kernel's security check during the \texttt{setresuid} syscall in order to set an unprivileged application's process ID to root. The success rates of each attack were 0.53\% (attack one), 0.63\% (two), and 0.41\% (three) following 22,118, 12,705 and 18,968 experiments respectively over 17--21 hours.

The NCC Group~\cite{ncc:group} (2020) published a voltage glitch vulnerability on a MediaTek MT8163V SoC (ARM Cortex-A53). A potential voltage glitch was discovered after the first bootloader is loaded from eMMC into RAM, causing a signature verification check to be skipped.  A glitch trigger was implemented on eMMC activity to dynamically insert and load an unauthorised boot component containing code for privileged execution (EL3) with a success rate of 15.21\%--23.44\%. While not practically demonstrated, the authors posited that the glitch could be used to load an unauthorised TEE image.

The previous attacks mounted voltage FIAs from external generators, but recent work has shown that dynamic voltage and frequency scaling (DVFS) frameworks can be used for on-SoC fault generation. DVFS regulates the operating frequency and voltage of CPUs for minimising its energy consumption. In 2019, Qiu et al.~\cite{qiu2019voltjockey} released the VoltJockey attack on TrustZone-based SoCs, which made use of kernel drivers for controlling the power management IC (PMIC) in software. After establishing root access, the CPU voltage was lowered to the point where cross-core faults were generated on a multi-core system (Figure \ref{fig:voltjockey}). Because both worlds of execution use the same physical processor, faults could be triggered in the non-secure world for key recovery from the secure world. The authors could induce byte errors during RSA signature verification and AES's eighth round using reference software implementations. The former was used to force the loading of an unsigned TEE application by the TrustZone OS. Proof-of-concept attacks were mounted on a Google Nexus 6 smartphone with a Qualcomm APQ8084AB SoC, with presented rates of 4.6\% (RSA verification) and 2.2\% (AES key recovery). The code for these attacks, however, have never been released. On non-mobile systems, similar attacks have been used since against Intel SGX on X86-64 platforms~\cite{voltpillager,qiu2019voltjockey1,murdock2020plundervolt}.

\subsubsection{Clock Glitch Attacks}
\label{sec:c_fi}

\begin{figure}
    \centering
    \includegraphics[width=\linewidth, interpolate=on]{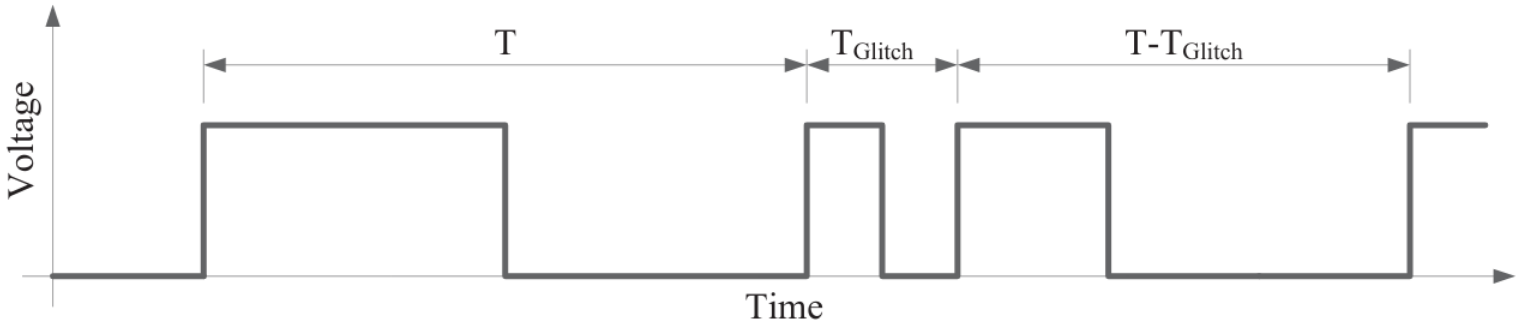}
    \caption{Using an FIA to introduce an additional positive clock edge~\cite{korak2014effects}.}
    \label{fig:clock_glitch_fig}
\end{figure}

Clock glitch attacks perturb device clock cycles, e.g.\ introducing additional edges (Figure \ref{fig:clock_glitch_fig}), in order to induce hardware synchronisation issues. This can trigger instruction misses, caused by forcing the execution of an instruction before the CPU has completed the previous one, and data misreads from attempting to read values before the memory has latched out the request. Clock glitches have attracted significant attention against simpler systems~\cite{balasch2011depth,markantonakis2009attacking,mayes2008smart}. However, at the time of publication, they have found limited application to mobile device SoCs.

An exception is \textsc{CLKScrew} by Tang et al.~\cite{tang2017clkscrew} (2017), which uses software-controlled power management for cross-core fault generation on TrustZone-enabled SoCs. The attack also exploits DVFS---see Qiu et al.~\cite{qiu2019voltjockey} in \S\ref{sec:v_fi}---for overclocking the CPU to fault secure world operations from the non-secure world. Two attacks are presented: 1), secure world AES key recovery; and 2), corrupting RSA signature verification checks used by the trusted OS prior to loading TAs.  \textsc{CLKScrew} has several prerequisites: kernel access is required to control the power manager, the core clock frequency must be modifiable, interrupts must be disabled, and the target TA must be repeatedly invoked from the non-secure world to decrypt arbitrary ciphertexts. The authors evaluated a Google Nexus 6 using a Qualcomm Krait-based SoC and a TEE OEM implementation. On average, useful faults were generated at a 5\% rate for inducing a one-byte fault to a desired AES round for AES key recovery, and 1.51\% for the RSA attack.

Besides \textsc{CLKScrew}, no other publications have used clock glitches against mobile SoCs to our knowledge. In the recent MCU literature, Korak and Hoefler~\cite{korak2014effects} (2014) used clock FIAs to skip arithmetical (\texttt{adds}), branch (\texttt{beq} and \texttt{breq}), and memory instructions (\texttt{ldr} and \texttt{str}) on a 16-bit AVR ATxmega256 and 32-bit ARM Cortex-M0.  They discovered that solely using clock glitches was ineffective on the Cortex-M0, leading to a combined approach of using voltage underfeeding in conjunction with a clock FIA; the ARxmega256 did not require underfeeding. The attacks principally affected the fetch and execute stages of the MCUs' instruction pipelines, which could be induced with high probability (up to 100\%). The test hardware was a Xilinx Spartan-6 XC6SLX45 FPGA with a NXP LPC 1114 (ARM Cortex-M0) and ATxmega256A3 extension board. No attacks were presented against security mechanisms or cryptosystems.

In 2014, Bl\"{o}mer et al.~\cite{blomer2014practical} presented a clock FIA against an Atmel AVR XMEGA A1 that executed an implementation of pairing-based cryptography (PBC) from the RELIC toolkit~\cite{relic}. The authors discovered a clock glitch vulnerability that triggered the skipping of a jump instruction (\texttt{rjmp}), which was used for secret key recovery following 4,000 fault attempts (0.025\% success rate).

\begin{figure}
    \centering
    \includegraphics[width=0.98\linewidth, interpolate=on]{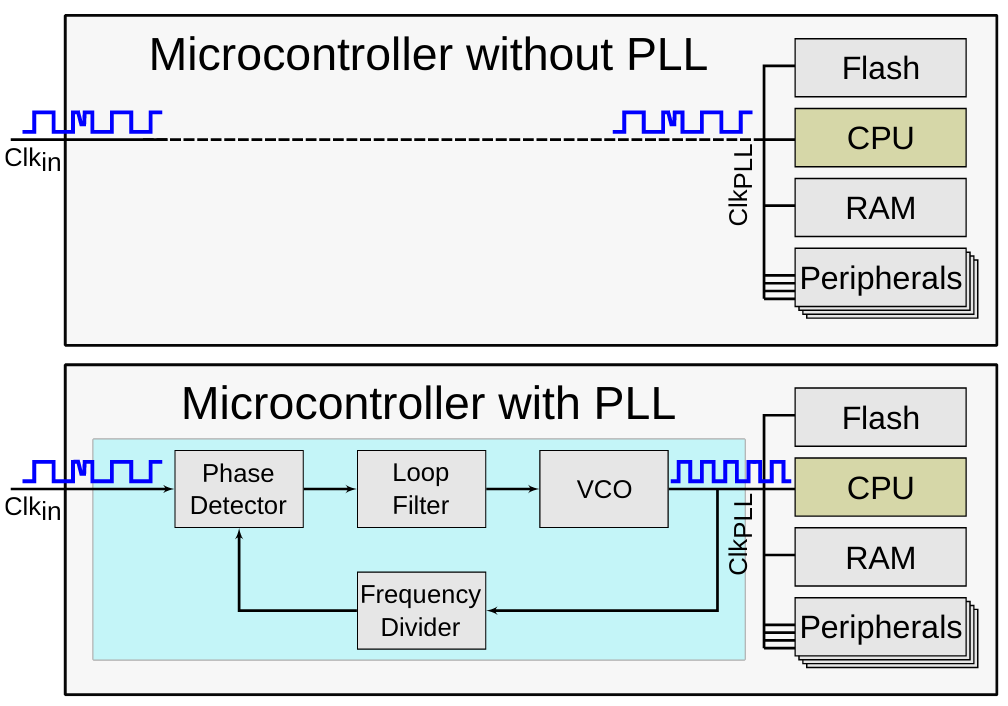}
    \caption{A microcontroller with and without a phase-locked loop (PLL) circuit using an external clock input~\cite{selmke2019peak}.}
    \label{fig:pll}
\end{figure}

A difficulty of performing clock glitches on modern systems is the prevalence of phase-locked loop (PLL) circuits (Figure \ref{fig:pll}), which process external clock sources into the system frequency thereby acting as a filter. Traditional FIAs that manipulate external clock frequency do not, therefore, straightforwardly influence the internal clock on modern ICs. In 2019, Selmke et al.~\cite{selmke2019peak} were able to successfully attack a PLL-equipped STM32F0308R with an ARM Cortex-M0 CPU. The approach necessitated an extreme clock glitch that effectively overclocked the PLL in order to produce deterministic perturbations in its output frequency. Using this, full key recovery was achieved against a reference software implementation of AES in electronic codebook (ECB) mode. In total, 1,000 faults were injected of which 16.4\% were exploitable.

\subsubsection{Heating Attacks}
\label{sec:heating}

Exposing ICs to extreme temperatures has been known for $>$15 years to generate multi-bit errors in DRAM memory and read/write threshold mismatches in non-volatile memory~\cite{kim2007faults}. Govindavajhala and Appel~\cite{govindavajhala2003using} (2003) presented a proof-of-concept showing that temperature-induced bit errors can lead to security vulnerabilities on a personal computer.  A 50W light bulb increased the DRAM working temperature to 100$^{\circ}$C, which triggered up to 10 flipped bits per 32-bit word with a 71.4\% probability. The authors used this to circumvent the Java type system to expose vulnerabilities in two commercial Java virtual machine (JVM) implementations. In general, heating attacks are among the most destructive of FIAs, leading to long-lasting component/IC damage if the exposure is too high and for too long~\cite{barenghi2012fault}.  As far as we are aware, no heating attacks have been published against mobile device SoCs.

In the MCU literature, Hutter and Schmidt~\cite{hutter2013temperature} (2013) published heating FIAs on an AVR ATmega162. A low-cost laboratory heating plate heated the MCU to $>$150$^{\circ}$C, beyond its maximum specified temperature limit (125$^{\circ}$C). The IC exhibited faults between a 152--158$^{\circ}$C window in which RSA-CRT decryptions were executed at 650ms intervals using a reference software implementation over a 70 minute period. This elicited 100 faults of which 31 were exploitable using the Boneh et al.~\cite{boneh1997importance} RSA-CRT fault attack to recover one of the prime moduli. In 2014, Korak et al.~\cite{korak} exposed an AVR ATmega162 to temperatures of up to 100$^{\circ}$C to facilitate clock glitches against an 8-bit smart card MCU. They showed that higher temperatures led to higher success rates for inducing erroneous instruction repetitions, replacements, and modifications of their destination registers. Attacks on particular cryptosystems or security mechanisms were not shown, however.

\subsubsection{Electromagnetic FI (EMFI)}
\label{sec:emfi}

Another widely studied attack vector is to expose device components to strong electromagnetic pulses. One of the earliest fault models was presented by Quisquater and Samyde~\cite{quisquater2002eddy} (2002) in which EM-induced Eddy currents in the target circuit are captured by its latches, thus generating bit faults. Subsequent fault models have been developed since, such as Raoult et al.~\cite{raoult2015electromagnetic} (2015) who examined the coupling of a near-field probe to a printed circuit board (PCB) micro-strip line. Generally, electromagnetic FIs (EMFIs) have greater spatial precision than other non-invasive FIAs. High-precision probes connected to EM pulse generators can be used to perturb specific IC regions while shielding other components. EMFIs have attracted significant attention from the research community since their use on smart cards in the early-2000s~\cite{quisquater2002eddy,markantonakis2009attacking,mayes2008smart}.

\paragraph{\textbf{EMFIs on Mobile Phone SoCs.}}
To our knowledge, Cui and Housley~\cite{cui2017badfet} (2017) presented the first attack on a moblie phone SoC, known as BADFET, which used second-order effects of EMFIs where faults in one component also induced faults in dependent components.  A low-cost test-bed (\$300 USD) was used to corrupt the contents of DRAM and NAND flash memory, which triggered a CPU instruction cache fault where bootloader code was loaded. The authors were able to skip into an unreachable code region of the bootloader containing a command line interface (CLI) for debugging purposes. Through this, a separate binary was loaded that exploited an existing vulnerability in the TrustZone SMC implementation (see \S\ref{sec:trustzone}) for privileged TEE code execution. A Cisco 8861 IP Phone with a Broadcom BCM11123 SoC was used as the DUT, where the attack could be repeated across 72/100 attempts (72\%).

In 2019, A{\"\i}t El Mehdi~\cite{ait2019analyzing} presented the first public results on applying EMFIs against a package-on-package (PoP) SoC. The initial research goal was to disable the Android lock-screen's timeout countermeasure to protect against brute force attacks. While this did not succeed, EMFIs could be used against the PoP SoC's upper DRAM package to trigger instruction corruption errors for modifying program control flow. Interestingly, the glitched instructions persisted in the CPU instruction cache as long as it was not replaced. The specific device and SoC models are redacted, however.

\begin{figure}
    \centering
    \includegraphics[width=0.87\linewidth,interpolate=on]{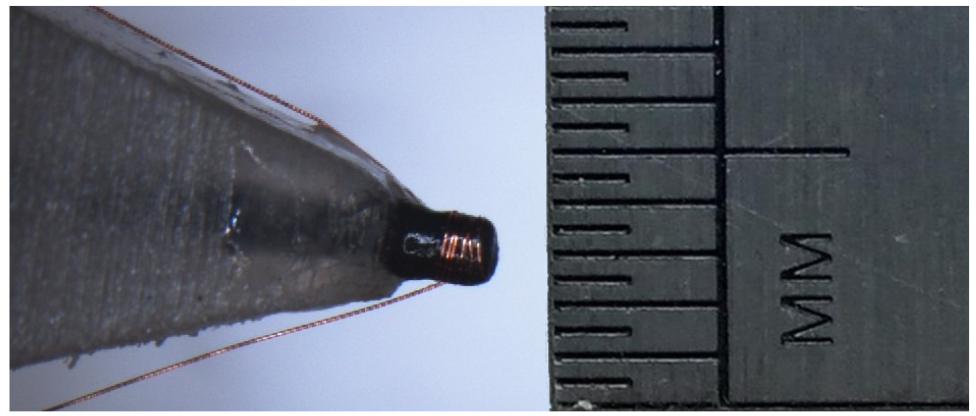}
    \caption{EMFI probe used by Gaine et al.~\cite{gaine2020emfi}.}
    \label{fig:gaine_probe}
\end{figure}

Gaine et al.~\cite{gaine2020emfi} (2020) presented a privilege escalation attack on a 64-bit (non-PoP) SoC with four ARM Cortex-A53 CPUs (1.2GHz) on a mobile development board, which ran version 4.14 of the Linux kernel from the Yocto Project's Sumo release. EM pulses were fired using an injection probe, shown in Figure~\ref{fig:gaine_probe}, on an XYZ motorised stage. Only one CPU on the DUT was sensitive to EMFIs after characterising the temporal and spatial requirements using a test program. An instruction skip vulnerability was identified with a conditional branch instruction (\texttt{cbz}) in the string comparison (\texttt{strcmp}) C function. This was exploited to bypass the password comparison check used by the substitute user command (\texttt{su}) using a software trigger in the \texttt{libpam} kernel module. Without DVFS enabled, EMFIs were generated every two minutes for the same CPU with a fixed frequency and probe position (overall success rate of 2\%).  With DVFS enabled, 21 of 6000 FIs were useful (0.35\%), equivalent to one success for every 300 attempts (15 minutes).

In other work, Trouchkine et al.~\cite{trouchkine2019fault} (2019) detailed micro-architectural and instruction-level software methods for expediting EMFI characterisation on modern SoCs. Using these, the authors were able to establish the location of useful faults on ARM and Intel CPUs; that is, whether they perturbed data in CPU registers, the pipeline, MMU, caches or external memory. Experimental results showed that faults on CPU registers, the CPU pipeline, or memory could be identified on a Broadcom BCM2837 SoC (ARM Cortex-A53 at 1.2GHz) and an undisclosed Intel Core i3 CPU with 95\% and 80\% accuracy respectively.

\paragraph{\textbf{EMFIs on MCUs.}}

Dehbaoui et al.~\cite{dehbaoui2013electromagnetic} (2013) mounted an EMFI-based key recovery attack on a reference software AES implementation on a 32-bit ARM Cortex-M3 MCU (24MHz). The work targeted the AES round counter at the ninth and tenth rounds to induce an additional round of execution with high success (up to 100\%). This enabled feasible cryptanalysis with which the encryption key could be recovered using two pairs of correct and faulty ciphertexts. 

Menu et al.~\cite{menu2019precise} (2019) presented data corruption EMFIs against the transfer bus from flash memory to the internal buffer of an Atmel SAM3X8E MCU (ARM Cortex-M3). Without disturbing code execution, the authors could reset data items of length 0--128 bits with byte-level precision by targeting the pre-fetch mechanism using an EM generator with a square voltage pulse (200V maximum amplitude), a 6ns minimum width and a 2ns transition time. Four components were shown to be vulnerable: the flash memory, the 128-bit pre-fetch buffer, bus interfaces, and the register file. A reference software AES implementation was then interrogated using a key is stored in flash memory. For the first attack, the EMFI could reset the value of the entire key as it was fetched from memory (100\% repeatability). The second applied the DFA from Biham and Shamir~\cite{biham1997differential}, which assumes the attacker can collect ciphertexts of known plaintexts while incrementally resetting one bit of the secret key until a zeroed key is reached. This was used to reduce the potential key-space of a 128-bit AES key to $16 \times 2^{8}$, which could be brute-forced. The final investigation presented a fault attack that uses an EMFI to induce a persistent fault in AES S-box look-up elements. Encryptions using the S-box produced faulty ciphertexts if the corrupted entry was accessed, where it was shown that the effective key entropy reduced to 32-bits when four such S-box elements were used.

Similar to the aforementioned work in \cite{trouchkine2019fault}, Maj\'{e}ric et al.~\cite{majeric2016electromagnetic} (2016) showed how EM side-channels could expedite EMFI parameter setting against an AES implementation aboard the security co-processor of an IoT SoC. The EMFI location was set using EM emissions corresponding to particular Hamming weights during intermediate AES operations. The ninth AES round was successfully faulted at a rate of 0.03\% after which faulty ciphertexts could be used for key recovery using the Piret and Quisquater~\cite{piret2003differential} DFA. Unfortunately, the identities of the SoC and its co-processor are not disclosed.

Liao and Gebotys~\cite{liao2019methodology} (2019) combined EMFIs and overclocking for disrupting the pre-fetch stage of an 8-bit MCU (Microchip PIC16F687). After performing backside decapsulation, the authors caused instruction replacement errors by inducing bit-level opcode faults within several test programs. An exclusive-or (\texttt{xorwf}) and move-literal (\texttt{movlw}) instruction were particularly vulnerable, which could be replaced with alternatives---e.g.\ \texttt{goto}, inclusive-or (\texttt{iorwf}) and bit-clear (\texttt{bcf})---with a frequency of 1.8\%--98.7\%. A key recovery attack was then shown against a reference AES software implementation requiring 222 EM pulses and 5.3 plaintexts on average. It is mentioned that \texttt{goto} instruction replacement errors may be used bypass boot-time authentication checks, but a concrete attack was not presented.

Other than examining cryptographic implementations, several works have applied EMFIs to trigger general instruction and data corruption faults. Moro et al.~\cite{moro2013electromagnetic} (2013) discovered instruction and data corruption vulnerabilities by targeting the data and instruction buses on an ARM Cortex-M3. EMFIs were used to: 1), modify the values of \texttt{ldr} instructions to alter data flow; 2), generate hardware exceptions to modify control flow; 3), replace memory store (\texttt{str}) instructions; and 4), modify CPU registers. Experimental results explored the effect of EMFI pulse amplitudes on the number and relative frequency of bit faults on data fetches from flash memory. 

In a similar vein, Riviere et al.~\cite{riviere2015high} (2015) interrogated the instruction cache of an ARM Cortex-M4 by targeting an EMFI during the updating of the CPU's pre-fetch queue buffer with 96\% success. Using custom assembly programs, four instructions could be skipped while the subsequent four instruction could be replayed. No security-critical systems were compromised, but the authors posit that the fault model could be used against CRT-based RSA, AES, and for privilege escalation on ARMv7-M targets.

Elmohr et al.~\cite{elmohr} (2020) discovered an EMFI-based instruction corruption fault during the execution stage of load (\texttt{ldr}) instructions on an NXP LPC1114 (ARM Cortex-M0). Using a test program, the errors appeared when a branch (\texttt{bne}) instruction was fetched at the same time. EMFIs were used to fault two, three, four and instructions at the same time with a frequency of 31\%, 13\%, 12\% and 12\% respectively. Like \cite{liao2019methodology}, the MCU must be decapsulated.

\section{Side-channel Attacks}
\label{chapter:sca}

Rather than inducing abnormal conditions into electronic components, physical side-channel attacks exploit observable phenomena emitted during their operation. This can reveal key material and other data if the phenomenon can be reliably associated with data or operations under execution. The origin of these attacks date back to the disclosure of the NSA TEMPEST programme~\cite{friedman1972tempest}, which described how vulnerable cryptographic implementations can produce certain electromagnetic characteristics depending on the input key and data. The area has since attracted tremendous attention from the research community, with associated vulnerabilities still plaguing today's systems~\cite{camurati2018screaming,gnad2019leaky,wang2019side,wang2020far,zhou2005side,mangard2008power}.

A standard SCA testing setup was shown in Figure~\ref{fig:attack_setup}. Here, an oscilloscope acquires measurements (traces) of the chosen phenomenon, e.g.\ EM emissions, using probes attached to sensitive DUT regions, which are then analysed by the investigator on an external computer. Like FIAs, the target may be instrumented with hardware or software triggers to collect precise measurements of particular operations of interest. In existing literature, two SCA approaches have been widely utilised:

\begin{itemize}
    \item \textbf{Power analysis}: Leverages time-series differences in the target's power consumption during sensitive procedures~\cite{gnad2019leaky,picek2018performance,schmidt2010side,heuser2012intelligent,park2018power,msgna2014precise}. 
    \item \textbf{EM analysis}: Exploits electromagnetic radiation signatures produced by the target~\cite{aboulkassimi2011electromagnetic,kenworthy2012mobile,nakano2014pre,montminy2013differential,balasch2015dpa,goller2015side,longo2015soc,belgarric2016side,bukasa2017trustzone,camurati2018screaming,alam2018one,wang2020far}.
\end{itemize}

Beyond these approaches, two alternative methods have also been explored in the wider literature, albeit with limited applicability to mobile systems:
\begin{itemize}
    \item \textbf{Acoustic cryptanalysis}: Capitalises on properties of sound waves, such as the emission of certain frequencies and their change over time~\cite{genkin2014rsa}.
    \item \textbf{Temperature analysis}: Uses differences in the target's temperature during its operation~\cite{hutter2013temperature}.
\end{itemize}

We observe that only EM-based attacks have been used extensively on mobile phone SoCs. The others have been employed predominantly against simpler SoCs and MCUs, which we discuss as supplementary information. 

\subsection{Power Analysis}

Power analysis exploits device power consumption measurements to discover information about sensitive code and data under execution.  Modern ICs contain millions of transistors that act as voltage switches. These are continually switched on and off during execution, causing voltage fluctuations that can be measured using commercially available testing equipment. The resulting measurements are analysed, whether through direct inspection or using statistical analysis, for inferring information about cryptographic keys and individual instructions under execution~\cite{msgna2014precise,markantonakis2009attacking,schmidt2010side}. 

Interestingly, very few power analysis attacks have been conducted on mobile devices. In one exception, Genkin et al.~\cite{genkin2016ecdsa} (2016) used a power tap on an Apple iPhone 4 USB charging port, requiring a USB pass-through adapter with a 0.33$\Omega$ resistor placed in series with the ground line. The authors connected the phone to a battery pack via the adapter and collected measurements using voltage changes over the resistor. Through analysis, five ECDSA scalar-by-point multiplication operations (using the same point and four different scalar values) could be distinguished using OpenSSL's NIST P-521 implementation. Details were not presented regarding its effectiveness for full recovery or its transferability to other devices. The second instance we are aware of, published by Lisovets et al.~\cite{lisovets2021iphone} (2021), is an additional investigation of an EM-based side-channel attack on an iPhone 4, which we discuss in \S\ref{sec:emsca}. 

In general, power analysis attacks on mobile SoCs face a myriad of challenges. Firstly, their density renders it difficult to capture reliable measurements, with varying power planes, voltage regulators, and landside decoupling capacitors compared to MCUs. Modifying mobile SoCs to access reliable power sources is not a trivial endeavour and risks serious damage to the device.  Secondly, attacks must account for potential interference from multi-core CPUs and multitasking OSs; the power consumption of specific operations of interest must be distinguished from aggregate measurements of the entire system. Furthermore, modern PoP designs require researchers to account for the power consumption of multiple packages in addition to the aforementioned complexities.  All of these aspects dramatically complicate power analysis attacks on mobile devices.

In the remainder of this section, we cover recent research targeting embedded systems, e.g.\ IoT SoCs, which may provide a gateway for analysing mobile SoCs in future work.

\subsubsection{Key Recovery}
\label{sec:keyrecovery}

Most key recovery attacks employ a common set of techniques for analysing power traces. \emph{Simple power analysis} (SPA) involves directly inspecting traces to deduce secret information when measurements can be mapped to certain data properties, e.g.\ distinct voltages corresponding to individual key bits. \emph{Differential power analysis} (DPA), an advanced technique presented by Kocher et al.~\cite{kocher1999differential}, uses correlations between the Hamming weight of intermediate cryptographic operations and measurement traces to determine the likelihood of particular key bytes. \emph{Template analysis} involves collecting a set of traces and labelling them with the corresponding cryptographic operation. Freshly measured traces are then classified by mapping it to its closest matching template from the set.

Over the last 20 years, SPAs, DPAs, and template analyses have been studied in great detail on smart cards, FPGAs and embedded systems~\cite{markantonakis2009attacking,mangard2008power,fan2010state,moradi2010lightweight}. Recent work has explored machine and deep learning methods for providing greater discriminative power using fewer traces over existing statistical methods.  Heuser and Zohner~\cite{heuser2012intelligent} (2012) explored the use of support vector machines (SVMs) for this purpose. Power traces were collected from an 8-bit AVR ATMega-256-1 MCU (8MHz) using a reference AES software implementation. The AES S-box was used as the profiling target in which the SVM predicted the Hamming weight from the measurement trace. The authors were able to recover the key using 20 traces in low-noise and up to 60 traces in high-noise environments. Beyond power analysis, traditional machine learning has also been used for side-channel analysis in other problem spaces, such as recovering information about physically unclonable functions (PUFs). We refer to Hettwer et al.~\cite{hettwer2019applications} for a detailed survey of this work.

Deep learning (DL) methods---neural networks with multiple hidden layers---are also being increasingly used in state-of-the-art attacks, which can capture complex, non-linear interactions between power traces and target cryptographic operations.  Maghrebi et al.~\cite{maghrebi2016breaking} (2016) presented some of the first results applied to AES key recovery by targeting AES's first-round S-box. The first attack focussed on an unprotected FPGA-based AES implementation, showing that key recovery can be achieved with 200 traces using a convolutional neural network (CNN). In the second, an unprotected reference software AES implementation was examined on a Chipwhisperer---a development board for side-channel analysis with an AVR ATMega328P MCU. Using an autoencoder neural network, the first four AES key bytes could be recovered with 20 traces. In the final experiment, a masked AES software implementation on the Chipwhisperer was investigated; in the best case, an autoencoder and CNN could recover the secret key using 500 and 1,000 traces respectively.

The application of CNNs was also explored by Picek et al.~\cite{picek2018performance} alongside simpler machine learning methods. Similarly, they used the Hamming weight model at first-round AES S-box operations. Protected AES implementations were examined using the DPAContestV2~\cite{clavier2014practical} dataset, comprising 50,000 traces from an Atmel AVR MCU. A 91.2\% test accuracy was achieved for correct Hamming weight classification using a CNN. Simpler methods---XGBoost, Na\"{i}ve Bayes, and Random Forests---were also used effectively, where the key was recovered in under 10 traces. Wang~\cite{wang2019side} (2019) conducted similar work using CNNs for key recovery against two AVR ATXmega128D4 MCUs with a 128-bit AES-ECB reference software implementation. Interestingly, the average number of required traces differed significantly between the boards (160 and 400 traces per MCU).

As an alternative to tapping into current-carrying wires, Schmidt et al.~\cite{schmidt2010side} (2010) used miniscule power fluctuations measured at exposed I/O pins. The authors targeted five devices running a reference 128-bit AES software implementation without side-channel countermeasures: an 8-bit Atmel ATMega163 and AT89S2853 MCU, an NXP LP2148 MCU with a 32-bit ARM ARM7TDMI-S CPU, a Virtex-II Pro XC2VP7 FPGA, and an ASIC. Plaintexts were over a serial interface to the targets and, during encryption, an oscilloscope recorded the voltage variations at an exposed I/O pin. DPAs were reportedly successful on all devices, but the required number of traces were not given.

Recently, Gnad et al.~\cite{gnad2019leaky} (2019) showed that crosstalk in analog signals from adjacent on-SoC digital components could be exploited for key recovery. The threat model is an on-device malicious program that wishes to learn information about secret data used by a parallel program on another CPU core.  The authors used the effect of noise fluctuations generated by digital subsystems that were modulated into (on-SoC) analog-to-digial (ADC) converter signals. A spy program accessed ADC measurements while the victim used a software AES implementation from \texttt{mbedTLS} to execute single 128-bit AES encryptions of plaintext messages. Using a range of IoT SoCs---a ESP32-devkitC, an STM32L475 IoT Node, and an STM32F407VG Discovery---correlation analysis attacks were mounted by targeting the final AES round and analysing 10 million ADC noise traces.

\subsubsection{Instruction Profiling on MCUs}

\begin{figure}
    \centering
    \includegraphics[width=\linewidth,interpolate=on]{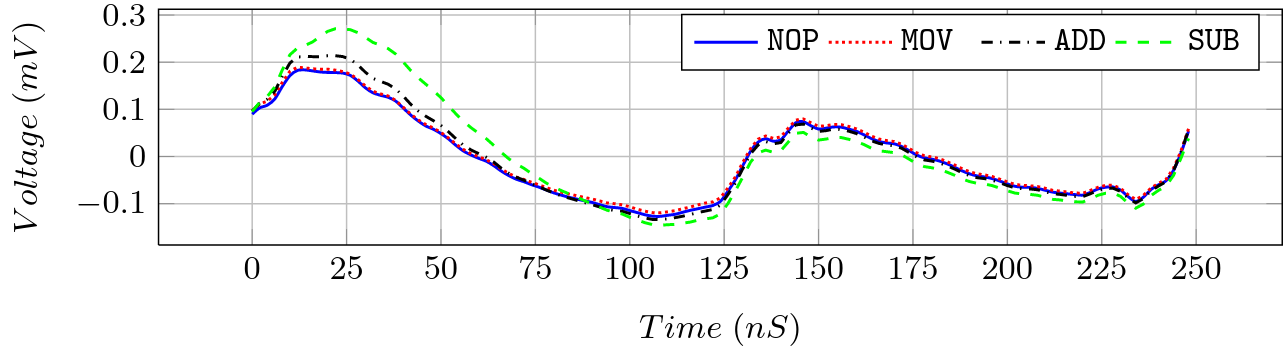}
    \caption{Power consumption of the \texttt{NOP}, \texttt{MOV}, \texttt{ADD}, and \texttt{SUB} ATMega163 instructions~\cite{msgna2014precise}.}
    \label{fig:msgna}
\end{figure}

Another less-studied use of power analysis is to identify individual instructions in use, which has applications to reverse engineering and detecting unauthorised program execution. In this area, Msgna et al.~\cite{msgna2014precise} (2014) analysed power consumption traces from an Atmel ATMega163 using the voltage drop across a shunt resistor on the MCU. 11 instructions were profiled---Figure \ref{fig:msgna} shows a sample trace of four ATMega163 instructions---using template analysis after applying dimensionality reduction as a feature extraction technique.   The work used two statistical models---k-Nearest Neighbour (kNN) and a multi-variate Gaussian probability distribution function---to classify instructions from their associated traces, achieving a 66.78--100\% recognition accuracy depending on the method. In 2018, Park et al.~\cite{park2018power} profiled a much larger set of instructions on an AVR ATMega328P MCU (112 in total). A classification model was developed based on quadratic discriminant analysis, which was trained on 2,500 power traces per instruction, achieving a recognition accuracy of 99.03\%.

\subsection{Electromagnetic Emission (EM) Analysis}
\label{sec:emsca}

 Using EM emissions to compromise vulnerable security systems can be dated back to 1943 when a Bell Telephone engineer discovered oscilloscope perturbations while using an encrypted teletype, the Bell 131-B2. The partial declassification of the NSA's TEMPEST programme in 1972 subsequently documented more elaborate efforts to subvert electronic systems in this way, as well as identifying potential countermeasures~\cite{wired:tempest,friedman1972tempest}. These side-effects arise from charge movements through transistor gates, wires etc., which emit electromagnetic radiation that can be measured using non-invasive means.\footnote{While EM-based SCAs are usually non-invasive, more reliable traces may be collected after invasive intervention, e.g.\ IC decapsulation, to eliminate unwanted attenuation.} EM analysis has been widely used to break security and cryptographic implementations on systems from simple smart cards to high-frequency SoCs~\cite{kocher1999differential,han2019side,ors2003power,messerges1999investigations,messerges2002examining}.  This section surveys state-of-the-art EM-based SCAs, which, unlike power analysis, has been employed extensively against mobile devices.

\paragraph{\textbf{EM SCAs on Mobile Phone SoCs.}}

\begin{figure}
    \centering
    \includegraphics[width=0.9\linewidth,interpolate=on]{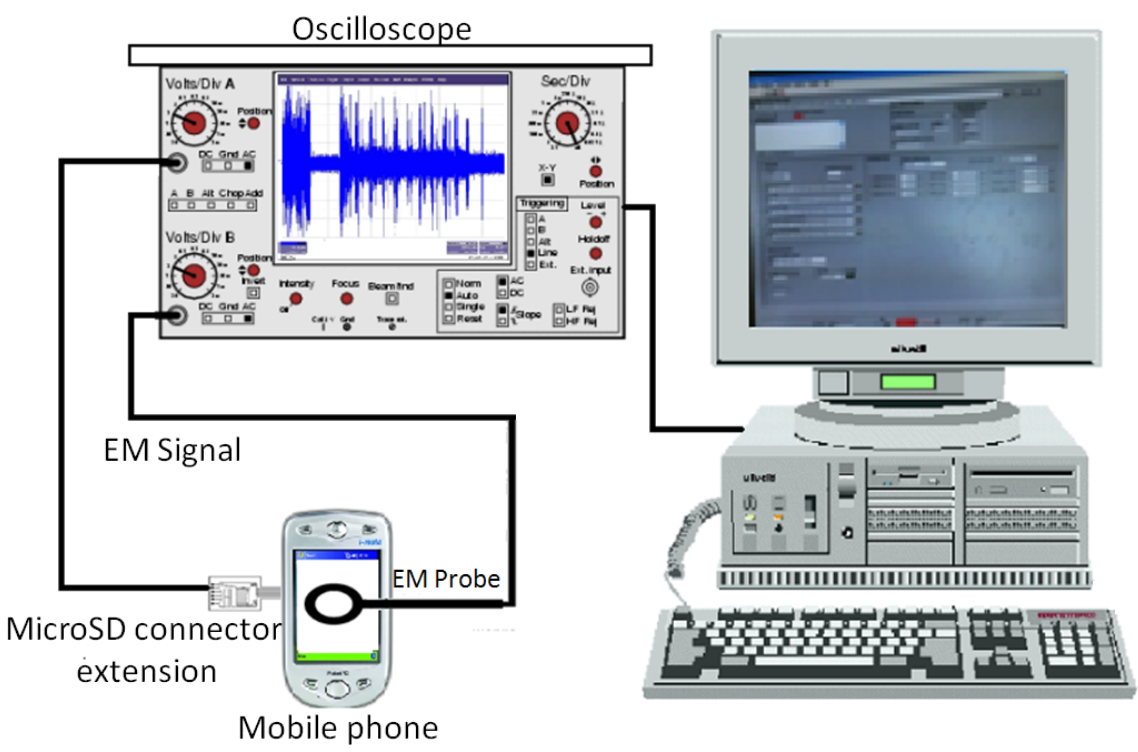}
    \caption{Experimental setup used by Aboulkassimi et al.~\cite{aboulkassimi2011electromagnetic}.}
    \label{fig:driss}
\end{figure}

To our knowledge, Aboulkassimi et al.~\cite{aboulkassimi2011electromagnetic} (2011) published the first EM-based SCA on a mobile phone, targeting reference AES software implementations on the Java Platform, Micro Edition (Java ME). Measurements were triggered using the device's microSD card interface, which were acquired using a commercially available EM probe and oscilloscope (see Figure \ref{fig:driss}). The authors presented methods to overcome temporal distortions in EM traces from just-in-time (JIT) compilation and garbage collection used by the Java Virtual Machine (JVM). Two methods---a spectral density-based approach (SDA) and a template-based resynchronisation approach (TRA)---were proposed to statistically navigate these issues. Using the second approach, one AES key byte could be recovered in an hour with 250 traces from an undisclosed device with a 32-bit RISC processor (370MHz). Like~\cite{barenghi2009low}, however, this platform can be considered obsolete nowadays.

In 2012, Kenworthy and Rohatgi~\cite{kenworthy2012mobile} reported ECC-, RSA- and AES-based key recovery attacks on three undisclosed mobile devices. The first---a \emph{``4G LTE smart phone from a major manufacturer''}---performed RSA-CRT encryptions (2048-bits) using a self-written square-and-multiply implementation. From a single trace, the secret exponent could be recovered using a Yagi antenna, magnetic probe, ICOM 7000 receiver, and an Ettus Research USRP digitizer costing \$1000 (USD) at the time.  The second device, \emph{``a mobile PDA,''} performed elliptic curve point multiplication over P-571 using a self-written double-and-sometimes-add implementation. The authors achieved full key recovery using a single trace from an EM probe located three meters (10 feet) from the device. The third, \emph{``another mobile phone from a major mobile manufacturer,''} used an OEM cryptographic library to perform 128-bit AES-CBC encryption on a 200kB data buffer. In this attack, traces corresponding to 12,500 individual AES block operations were required to perform DPA-based key recovery. For all targets, the SoCs were not disclosed nor were the use of any triggers.

Nakano et al.~\cite{nakano2014pre} (2014) presented EM-based key recovery attacks against ECC and RSA implementations from Android's Java Cryptography Extension (JCE) library on an undisclosed smartphone (832MHz clock frequency). The RSA implementation used a square-and-multiply approach, which was already known to be vulnerable to side-channel analysis. Like \cite{kenworthy2012mobile}, identifiable ECC double-and-multiply operations were also exploited. Both methods used simple power analysis at the 10 MHz and 20 MHz frequencies for key recovery using a single trace. The precise nature in which triggers were implemented are not disclosed.

In 2015, Balasch et al.~\cite{balasch2015dpa} presented EM-based DPA attacks against a Texas Instruments AM3358 Sitara SoC on a Beaglebone Black single-board computer (SBC) with an ARM Cortex-A8 (1GHz) and a Debian 7-based Linux distribution (kernel v3.8.13-bone47). The authors targeted an unprotected 128-bit AES software implementation and a bit-sliced version from K\"{o}nighofer~\cite{konighofer2008fast} (2008) designed to resist side-channel analysis. After analysis, 1.2 million traces were needed for key recovery using the bit-sliced implementation and a first-order DPA, while a second-order DPA was possible using 400,000 traces. In contrast, only 10,000 traces were required for the unprotected algorithm.
\begin{figure}
    \centering
    \includegraphics[width=\linewidth,interpolate=on]{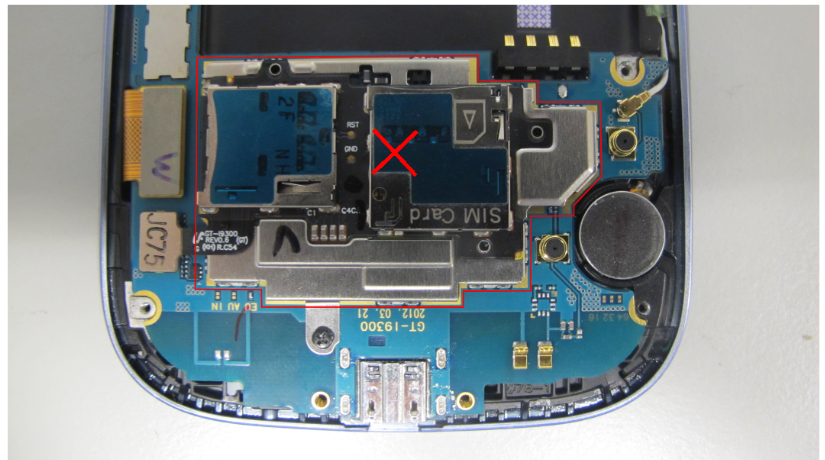}
    \caption{Device under test used by Goller and Sigl~\cite{goller2015side}. The red cross indicates the optimal EM probe position.}
    \label{fig:goller}
\end{figure}
\begin{figure}
    \centering
    \includegraphics[width=0.95\linewidth,interpolate=on]{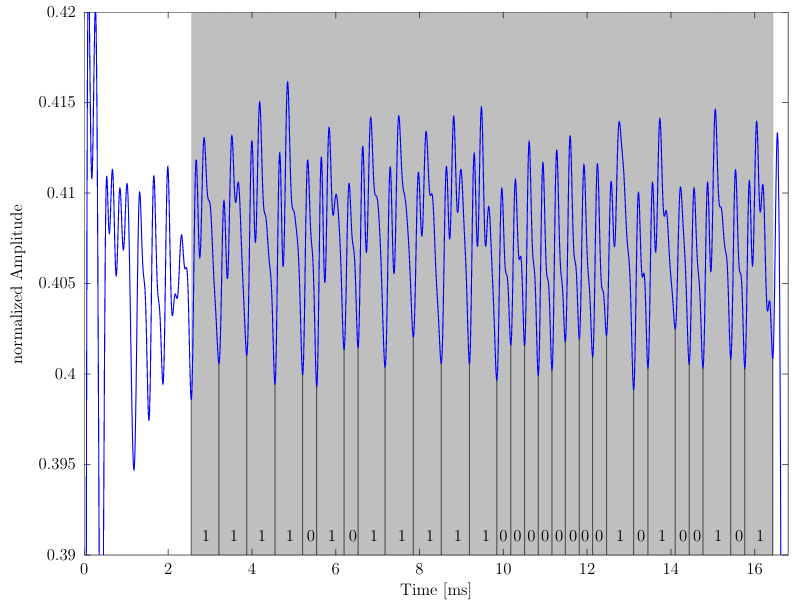}
    \caption{Average of 1063 RSA EM traces. Grey regions show the key bits recovered by Goller and Sigl~\cite{goller2015side}.}
    \label{fig:goller1}
\end{figure}

Goller and Sigl~\cite{goller2015side} (2015) attacked a reference software RSA implementation using the square-and-multiply method on five unnamed smartphones.\footnote{While formally undisclosed, a PCB serial number pictured in the paper corresponds to a Samsung Galaxy S3 (Figure \ref{fig:goller}).} The device's shielding plate was removed to reduce EM attentuation; a software-defined radio, a probe and a high-gain amplifier at a capacitor near the main CPU were then used for the measurement acquisition. Using simple power analysis, the waveforms were strongly correlated with individual bits of the secret key (Figure \ref{fig:goller1}). 276 traces were required for full key recovery with high confidence (0.999 correlation) and the shielding plate installed, while 170 were needed without the shielding.

Also using a BeagleBone Black SBC, Longo et al.~\cite{longo2015soc} (2015) conducted EM DPAs against two AES implementations: 1), software-based 128-bit AES in CBC mode taken from OpenSSL; and 2), a 256-bit AES implementation on the SoC's cryptographic co-processor. For the software implementation, 20,000 traces produced the correct key hypothesis, while the co-processor required 500,000 sets of 1,000 averaged traces (500 million traces in total).  Notably, they state that the latter may be useful against full-disk encryption implementation. While no exploits were presented to this end, it is one of the few papers to successfully attack a physical cryptographic co-processor, although no proof-of-concept code has been publicly released.

In 2016, Belgarric et al.~\cite{belgarric2016side} used EM analysis to identify ECC addition and multiplication operations in Android's Bouncy Castle ECDSA implementation. Full key recovery was performed using a lattice attack against an undisclosed smartphone with a Qualcomm MSM7225 SoC. After opening the device's external casing, an EM probe was placed on the SoC while traces were triggered using the USB interface. 39 ECDSA signature traces were needed for full key recovery, taking 102 seconds; the authors were also able to recover an Android Bitcoin wallet key as a example use case. Similar work was published concurrently by Genkin et al.~\cite{genkin2016ecdsa} (2016) who achieved full key recovery against OpenSSL's ECDSA implementation on iOS and Android. The attack was less invasive than \cite{belgarric2016side}, requiring only a probe to be placed in proximity of the device with no hardware or software triggers. A Sony-Ericsson Xperia X10 and iPhone 3GS were exploited using 5,000 signature traces from both devices, two of which (0.04\%) were useful.

In 2018, Alam et al.~\cite{alam2018one} presented the One\&Done attack for recovering RSA keys from a single decryption EM trace using OpenSSL (v1.1.0g). The attack, based on modelling potential control flow transitions of Montgomery multiplications, examined emissions at 40MHz around the target device's clock frequency using an Ettus USRP B200-mini software-defined radio. Interestingly, the authors subverted a side-channel resistant 2048-bit fixed-window constant-time RSA implementation and after the plaintext was blinded. Two Android phones were evaluated: a Samsung Galaxy Centura SCH-S738C with a Qualcomm MSM7625A (ARM Cortex-A5 at 800MHz) and an Alcatel Ideal with a Qualcomm Snapdragon 210 MSM8909 (ARM Cortex-A7). A OLinuXino SBC was also evaluated with an Allwinner 13 SoC (ARM Cortex-A7). During experimental validation, 95.7\%--99.6\% of the target key bits could be recovered depending on the chosen platform.

To our knowledge, Bukasa et al.~\cite{bukasa2017trustzone} (2017) were the first to investigate the EM properties of program execution in ARM TrustZone. They analysed an unprotected reference AES software implementations and a mock PIN verification algorithm on a Raspberry Pi 2 with a Broadcom BCM2836 SoC (quad-core ARM Cortex-A7 at 900MHz). The effect of multi-core \emph{vs.}\ single-core and secure world \emph{vs.}\ non-secure world execution was examined. Using template analysis, key recovery could be achieved by targeting AES's first-round S-box and collecting 150,000 EM traces. The key could be recovered with a success rate of 17.81\%--38.30\% depending on the system configuration. Multi-core execution in the secure world yielded the lowest success (17.81\%), while single-core execution in the non-secure world with the MMU disabled produced the best results (38.30\%). Leignac et al.~\cite{leignac2019comparison} (2019) further examined TEE \emph{vs.}\ REE execution using EM emissions from a reference software AES implementation on a HiKey SBC with an eight-core ARM Cortex-A53 CPU (1.2GHz), Android Oreo in the REE, and Trustonic Kinibi as the TEE OS. This same SoC is used by the Huawei P88 and Honor 5A smartphones. They showed that the TEE was still vulnerable, although synchronising the correlation power analysis attack was more complicated. TEE AES key recovery was achieved after 6,000 traces, while the REE required 10,000 to produce the correct key hypothesis.

\begin{figure}
    \centering
    \includegraphics[width=\linewidth,interpolate=on]{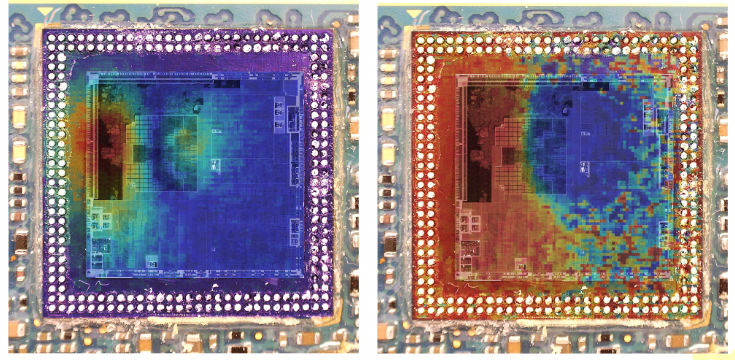}
    \caption{Lock-in thermography of the SoC die used by Vasselle et al.~\cite{vasselle2019breaking} showing heat maps of IR amplitudes (left) and in-phase AES emissions (right).}
    \label{fig:vasselle}
\end{figure}

In 2019, Vasselle et al.~\cite{vasselle2019breaking} published the first publicly available side-channel attack on a PoP SoC, which recovered a decryption key used by on-SoC boot ROM for decrypting BL1 firmware during the secure boot process (see \S\ref{sec:secure_boot}). The main SoC was first separated from the PoP DRAM package using a hot air station, as the DRAM package produced the most EM interference. This process required practice on multiple PoP units to prevent irreparable damage to the main SoC. EM measurements were acquired by targeting the on-SoC cryptographic co-processor using a Lecroy WaveRunner oscilloscope, a Langer H-Field probe, and a low-noise amplifier. Lock-in thermography was used to identify the co-processor from infrared measurements during encryption and decryption operations (Figure \ref{fig:vasselle}), which was used for accurately positioning the EM probe. 2,500 traces of 5,000 firmware binary decryptions were acquired (12.5 million AES-CBC operations in total). Correlation power analysis was then used to recover the co-processor's AES key from two firmware encryptions and 2,500 averaged traces. All details pertaining to the identity of the PoP, SoC and cryptographic co-processor are redacted; information about the trigger implementation is also not made public. We note that the approach worked only for BL1 as the missing PoP DRAM package caused a fatal boot deadlock at BL2.

\begin{figure}
    \centering
    \includegraphics[width=0.9\linewidth,interpolate=on]{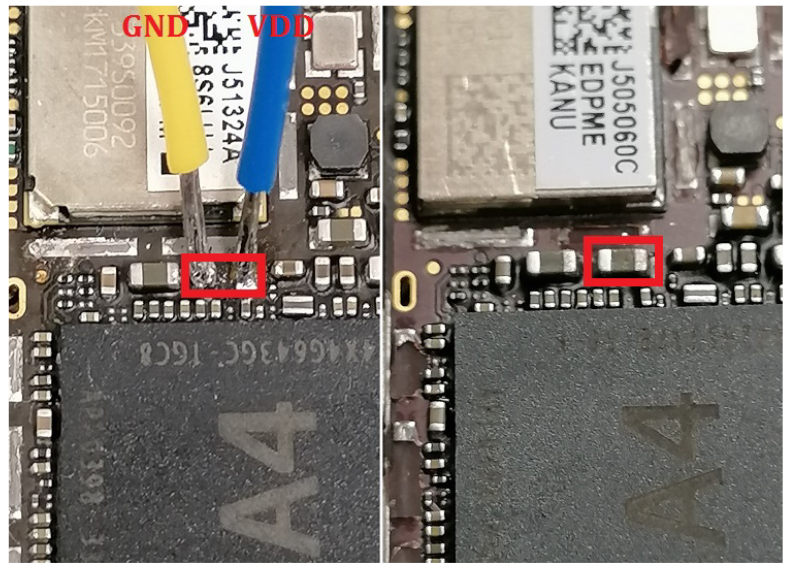}
    \caption{Partial PCB modifications made by Lisovets et al.~\cite{lisovets2021iphone} for providing an artificial power supply to the Apple A4 via an external board (modified: left; unmodified: right).}
    \label{fig:pcb-lisovets}
\end{figure}
Recently, Lisovets et al.~\cite{lisovets2021iphone} (2021) described the first physical SCA on an Apple iPhone 4 for extracting the 256-bit hardware-based user identifier (UID) key. The UID is used with the user's passcode to generate a passcode key from a password-based KDF to unwrap keys from the system keybag, which contains per-file data encryption keys for FBE using the SoC's AES encryption engine.\footnote{The work targets an Apple iPhone 4, which does not include a Secure Enclave Processor found on modern iPhone models (\S\ref{sec:proprietary}). The SEP prevents the application processor from accessing the UID for ultimately unwrapping FBE keys. Exploiting vulnerable boot ROM is, therefore, not enough to access the AES engine to mount the same attack.} A two-part attack was conducted: 1), using a known bootloader exploit to introduce an unauthorised component for submitting unlimited encryption queries to the AES engine; which was followed by, 2), using EM correlation power analysis against the AES engine. EM traces were measured from a probe on the Apple A4 SoC using a trigger from a GPIO pin re-configured by the unauthorised bootloader. For the specific equipment, a Langer EMV-Technik RF-B 0,3-3 EM probe, a Langer EMV-Technik PA 303 SMA amplifier, and a LeCroy WaveRunner 8254M oscilloscope (2.5GHz bandwidth, 40GS/s sampling rate) were used for trace acquisition.  Key recovery was achieved after 300 million traces collected over a two week period, which required up to three hours of analysis using two Nvidia RTX 2080 TI GPUs. In addition to the EM SCA, a power analysis attack was also mounted by measuring the voltage across a $1\Omega$ shunt resistor on an external board connected between the external power supply and the iPhone. This required invasive modifications to the PCB (Figure \ref{fig:pcb-lisovets}). The UID could then be recovered after 200 million traces using correlation analysis.  The authors state that the attack could be extended to modern models, although SCA countermeasures are likely to be present.

\paragraph{\textbf{SCAs on MCUs and Related Systems.}}

In the broader literature, Montminy et al.~\cite{montminy2013differential} (2013) extracted AES keys using a reference 128-bit AES-ECB software implementation on a Stellaris LM4F232H5QD MCU with a 32-bit ARM Cortex-M4F (50MHz). The authors targeted intermediate operations during the first AES round using the Hamming weight attack model. EM traces were acquired using a near-field probe and a software-defined radio to collect EM waveforms using a modified digital television receiver that costed \$20 (USD) at the time. A correlation analysis attack was then mounted to extract all of the AES key bits using 100,000 traces without instrumenting any triggers.

On an Intel SBC, Saab et al.~\cite{saab2016side} (2016) presented an EM DPA attack against the Intel AES-NI cryptographic instruction set extension on an Intel Core i7-3517UE. A dedicated C application was developed that made called the Intel AES-NI sample library (version 1.2), which looped over AES-256 (CBC mode) calls to the \texttt{iEnc256\_CBC} assembly routine. After collecting 1.5 million traces over 17 days, statistically significant Hamming weight leakage was discovered arising from CPU cache loading differences. A second investigation also discovered Hamming weight information leakage caused by the mixing of round keys between successive AES rounds, which was found from $\sim$1.3 million traces collected over 22 days. The authors provide initial evidence that Intel's AES-NI is not well-protected against side-channel analysis. This was further reinforced in the PLATYPUS attack by Lipp et al.~\cite{lipp2021platypus} (2021), who demonstrated a software-based power analysis attack using the Intel Running Average Power Limit (RAPL) interface, which leaked AES-NI key information used by Intel SGX enclaves.

In 2018, Camurati et al.~\cite{camurati2018screaming} showed that mixed-signal SoCs can leak exploitable EM emissions from the effects of analog/digital component crosstalk. Specifically, this occurred when the CPU noise was modulated into the radio transceiver's analog emissions. The authors investigated a Nordic Semiconductor nRF52832, a Bluetooth SoC with a 2.4GHz transceiver and an ARM Cortex-M4 CPU; and a Qualcomm Atheros AR9271, a wireless 802.11N SoC. 128-bit AES keys could be recovered at a 10 meter distance by observing EM emissions produced by \texttt{tinyAES} and \texttt{mbedTLS} in the 2.4GHz spectrum. Using template analysis, key recovery was demonstrated against \texttt{tinyAES} at a 10m distance using 70,000 traces for offline template creation and 428 traces for the actual attack For \texttt{mbedTLS}, a successful attack was mounted at a 1 meter distance with 40,000 traces.
 
In 2020, Benadjila et al.~\cite{benadjila2020deep} published work on using deep learning for EM analysis. The authors comprehensively assessed the application of CNNs to perform AES key recovery using the Hamming weight model. They targeted masked and unmasked reference software AES implementations on an 8-bit AVR microcontroller (ATmega8515) and released a publicly available dataset, the ASCAD dataset, to facilitate future research. Surprisingly, it was found that the VGG-16 image recognition network architecture was effective for EM-based side-channel analysis.

On an IoT SoC, Wang et al.~\cite{wang2020far} (2020) also applied deep learning for recovering AES keys using EM emissions. Three neural networks were trained---two CNNs with different layer configurations and one multi-layer perceptron (MLP) network---on traces captured from five Bluetooth devices at five distances from the target. Like \cite{camurati2018screaming}, the Nordic Semiconductor nRF52832 SoC was used with a reference 128-bit AES software implementation from the \texttt{tinyAES} C library. 500,000 traces were collected across various distances from the device (wired, then at 1m--8m) and in different locations (an office and corridor environment). In the best case, 367 traces of the same encryption were needed to recover an AES sub-key at a 15m distance using a 24dBi TP-Link TL-ANT2424B antenna and a software-defined radio.

\subsection{Other Physical Side Channels}

Another side-channel explored in historical literature is \emph{acoustic cryptanalysis}, although it has not been applied successfully to mobile SoCs to the best of our knowledge. This involves analysing sound waves produced by a system whose emitted pitch is dependent on the current operation. In 2014, Genkin et al.~\cite{genkin2014rsa} demonstrated that RSA key extraction using acoustic cryptanalysis was feasible against a Lenovo laptop from a smartphone in its proximity. The authors targeted GnuPG's RSA implementation using a laboratory microphone setup and a Samsung Note II, showing that a 4096-bit key could be recovered within one hour using audible and ultrasonic sound emanations.  

Examining \emph{temperature emissions} is also a known vector for side-channel analysis that, likewise, has found no application to mobile SoCs.  In 2013, Hutter and Schmidt~\cite{hutter2013temperature} demonstrated information leakage from an 8-bit AVR ATmega162 MCU by monitoring its temperature using a PT100 sensor element with a 100ms thermal response time.  As a proof of concept, the move (\texttt{mov}) instruction was used to move all possible values of one input byte (i.e.\ 256) into 24 internal registers. The temperature was measured for a period of 20 seconds, where noticeable increases were apparent depending on the Hamming weight of the processed value; however, a full key recovery attack was not shown. Crucially, the attack is appropriate only for simple devices that use long-running operations with low-frequency signatures on a single thread of execution. As such, it is unlikely to be practical against modern mobile systems.

\section{Evaluation}
\label{sec:eval}

This section evaluates state-of-the-art FIAs and SCAs, compares their features and results, and identifies challenges and limitations arising from existing literature.

\subsection{Comparison Criteria}

The previous sections showed that a large range of techniques, platforms, algorithms, and success rates have been reported by existing fault injection and side-channel attacks. To assist in their comparison, we distilled existing research into a common set of characteristics for evaluating their prerequisites, goals and results. These are as follows:
\begin{itemize}
    \item \textbf{Work} and \textbf{year}: The reference under comparison and its year of publication.\footnote{We note that older attacks may generalise poorly to today's devices if obsolete platforms were evaluated, e.g.\ due to the use of modern system-on-chip design features and the deployment of countermeasures.}
    \item \textbf{Attack class}: The type of SCA or FIA that was used, such as a voltage or clock glitch, EMFI, or power analysis attack.
    \item \textbf{SoC type}: The high-level SoC class, i.e.\ single- or multi-core, mixed-signal and/or high-frequency SoC. If a non-mobile phone SoC was used, such as an IoT SoC or MCU, then this is stated explicitly.
    \item \textbf{Trigger?}: Whether the DUT was instrumented with hardware/software triggers to generate precise faults or acquire side-channel measurements. 
    \item \textbf{Evaluation platform}: The device, SoC, MCU, or microprocessor model used as the target of attack.
    \item \textbf{Attack prerequisites}: Any preconditions that must be satisfied before their execution, such as gaining kernel-level code execution, decapsulating the SoC, or the use of a specific microprocessor architecture.
    \item \textbf{Implementation}: The software or hardware implementation type targeted by the SCA or FIA; for example, a proprietary OEM, an open-source, or a reference implementation. Generally, we consider attacks on OEM targets to most likely to generalise to other commercial implementations.
    \item \textbf{Success criteria}: The reported number of faults/traces required to mount the attack and/or the accuracy of the proposed method (if reported).
\end{itemize}

Lastly, existing work covers a multitude of scenarios, such as AES key recovery, privilege escalation, and skipping verification checks during secure boot processes. We abstracted these attack goals into 10 distinct categories to help researchers gauge their potential applicability:

\begin{enumerate}
    \item \label{item:sw_priva} \textbf{TEE code execution}: Enables privileged access to different TEE targets on the device, including TEE secure world applications in lower protection levels, device drivers to security-critical peripherals, and TEE memory management.
    
    \item \label{item:sw_kr}  \textbf{TEE AES/RSA/ECC key recovery}: Recover cryptographic keys from services that use AES-, RSA- or elliptic curve-based cryptography in the secure world. This has potential applications to TEE key management systems (\S\ref{sec:gm_ta}) and FDE and FBE implementations (\S\ref{sec:fde}).
    
    \item \label{item:load_tas} \textbf{Load unauthorised TEE TAs}: Load unauthorised trusted applications, which may be used to further understand the TEE's internals and for attempting secure world privilege escalation attacks.
    
    \item \label{item:bypass_secureboot_v} \textbf{Bypass secure boot verification}: Circumvent boot-time procedures that enforce the loading of authorised components, e.g.\ OEM-signed bootloaders, with the aim of loading unauthorised self-signed or unsigned images.
    
    \item \label{item:nsw_pa} \textbf{REE code execution}: Gain privileged access to the REE, including device drivers, memory management, and applications in less privileged protection modes.
    \item \label{item:aes_kr} \textbf{AES/RSA/ECC key recovery}: Recover keys against AES-, RSA- or elliptic curve-based cryptography implementations in the non-secure world. 
    \item \label{item:bypass_sig} \textbf{Bypass verification checks}: Bypass run-time security verification steps, such as those using RSA- and ECC-based digital signature verification algorithms.
    \item \label{item:load_unauthorised_data_into_memory} \textbf{Load unauthorised data into RAM}: Corrupt the device's data flow integrity protections to load unauthorised data items, e.g.\ cryptographic keys, into RAM during program execution.
    \item  \label{item:key_reset} \textbf{Key resetting}: Perform run-time bit reset attacks on part or all of a cryptographic key to disrupt target cryptosystems.
    \item \label{item:re} \textbf{Reverse engineering}: Recover instruction-level information about device programs under execution.
\end{enumerate}

We mapped this set of criteria to each publication surveyed in this work. Table \ref{tab:comparative_fi} provides a comprehensive comparative summary of fault injection attacks from \S\ref{chapter:fi}, while a separate comparison is provided in Table \ref{tab:comparative-sca} for side-channel attacks from \S\ref{chapter:sca}.

\begin{sidewaystable*}
\begin{adjustbox}{width=\columnwidth}
\begin{threeparttable}
\caption{Summary of recent fault injection attacks on mobile and embedded systems.}
\label{tab:comparative_fi}
\def\arraystretch{1.5}
\begin{tabular}{|c|c|c|c|c|c|c|c|c|c|}
\hline
\rowcolor{black!85}
 \textcolor{white}{\textbf{Work}} &  \textcolor{white}{\textbf{Year}} & \textcolor{white}{\textbf{Class}} & \textcolor{white}{\textbf{SoC Type}} & \textcolor{white}{\textbf{Evaluation Platform}} & \textcolor{white}{\textbf{Prerequisites}} &  \textcolor{white}{\textbf{Implementation}} & \textcolor{white}{\textbf{Trigger?}}  & \textcolor{white}{\textbf{Success Rate/Criteria}} &  \textcolor{white}{\textbf{Attack Goals}} \\\hline
 
 Vasselle et al.~\cite{vasselle} & 2018 & \multirow{2}{*}{LFI} & Multi-core & \makecell{ARM Cortex-A9} & \multirow{2}{*}{SoC decapsulation} & \multirow{2}{*}{Reference} & Y & Up to 95\% & \makecell{Secure world privileged access.}\\\cline{1-2} \cline{4-5} \cline{8-10}

 \cellcolor{gray!20} Colombier et al.~\cite{colombier2019laser} & 2019 & & MCU & \makecell{ARM Cortex-M3} &  &  & Y & Up to 100\% & \makecell{AES key recovery.}\\\hline
 
 %
 %
 Barenghi et al.~\cite{barenghi2009low} & 2009 & \multirow{8}{*}{VFI}  & \multirowcell{2}[-5pt]{Single-core} & \multirowcell{2}[-5pt]{ARM ARM926EJ-S} & \multirowcell{2}[-5pt]{---} & \multirowcell{2}[-5pt]{OpenSSL (v0.9.1i)\\ and Reference} & Y &\makecell{6.6\%--39\% (RSA-CRT attack)\\6.42\%--62.77\% (e-th Root)} & RSA key recovery.\\\cline{1-2} \cline{8-10}
 
Barenghi et al.~\cite{barenghi2010low} & 2010 &  & & & &  &  Y & \makecell{100K ciphertexts (different plaintexts)\\2M ciphertexts (same plaintext)} & AES key recovery.\\\cline{1-2} \cline{4-10}
 
 Timmers et al.~\cite{controlling_pc_fi} & 2016 &  & \multirowcell{2}[-5pt]{Multi-core} & \makecell{ARM Cortex-A9\\Xilinx Zynq-7010 SoC} &  AArch32 architecture & Reference & Y & 0.01\%--0.27\% & \makecell{Bypass secure boot verification.\\Load unauthorised bootloaders.\\Secure world privileged access.}\\\cline{1-2} \cline{5-10}
 
 Timmers \& Mune \cite{timmers2017escalating} & 2017 &  & & \makecell{ARM Cortex-A9} & \makecell{AArch32 architecture\\User-space access} & Linux kernel & Y & 0.41\%--0.63\% & Linux privileged access (REE).\\\cline{1-2} \cline{4-10}

 VoltJockey \cite{qiu2019voltjockey} & 2019 &  & \multirowcell{2}{\makecell{Multi-core\\1GHz+}}  &\makecell{Qualcomm APQ8084AB SoC\\Google Nexus 6} & \makecell{SW-controlled PMIC\\Kernel-space access} & \makecell{OEM TrustZone\\(AES and RSA)} & Y & \makecell{2.2\% (AES key recovery)\\4.6\% (RSA decryption fault)} & \makecell{Secure world AES key recovery.\\Load unauthorised TEE TAs.}\\\cline{1-2} \cline{5-10}
 
 NCC Group \cite{ncc:group} & 2020 &  &   & \makecell{ARM Cortex-A53\\MediaTek MT8163V SoC} & \multirow{4}{*}{---} & OEM & N & \makecell{15.21\%--23.44\%} & \makecell{Bypass secure boot verification.\\Load unauthorised bootloaders.}\\\cline{1-4} \cline{5-5} \cline{7-10}
 
 %
 %
 \cellcolor{gray!20} Korak \& Hoefler \cite{korak2014effects} & 2014 & \multirow{6}{*}{CFI} & \multirowcell{2}[-3pt]{MCU} & \makecell{AVR ATxmega256\\ARM Cortex-M0\\NXP LPC 1114} & & Reference & Y & Up to 100\% & \makecell{Bypass verification checks.\\AES key recovery.}\\\cline{1-2}\cline{5-5}\cline{7-10}
 
 \cellcolor{gray!20} Bl\"{o}mer et al.~\cite{blomer2014practical} & 2014 &  &  & \makecell{Atmel XMEGA A1} &   & RELIC toolkit~\cite{relic} & Y & \makecell{0.025\%} & \makecell{ECC (PBC) key recovery.}\\\cline{1-2} \cline{4-10}

 \textsc{CLKScrew}~\cite{tang2017clkscrew} & 2017 &  & \makecell{Multi-core\\1GHz+}  & \makecell{Qualcomm APQ8084AB SoC\\Google Nexus 6} & \makecell{SW-controlled PMIC\\Kernel-space access\\Repeated TEE TA invocation} & \makecell{OEM TrustZone\\(AES and RSA)} & Y & \makecell{5\% (AES key recovery)\\1.51\% (RSA verification fault)} & \makecell{Secure world AES key recovery.\\Load unauthorised TEE TAs.}\\\cline{1-2} \cline{4-10}

 \cellcolor{gray!20} Selmke et al.~\cite{selmke2019peak} & 2019 &  & \multirowcell{7}[-3pt]{MCU} & \makecell{ARM Cortex-M0\\STM STM32F0308R} & \multirowcell{7}[-3pt]{---}   & \multirowcell{7}[-3pt]{Reference} & Y & \makecell{16.4\%} & \makecell{AES key recovery.\\Glitch PLL-equipped CPUs.}\\\cline{1-3} \cline{5-5}\cline{8-10}

 \cellcolor{gray!20} Hutter \& Schmidt~\cite{hutter2013temperature} & 2013 &  \multirowcell{2}[-3pt]{HFI +\\CFI} &  &  \multirowcell{2}[-3pt]{AVR ATmega162I}&  &  & N & \makecell{31\%} & \makecell{RSA key recovery.}\\\cline{1-2} \cline{8-10}

 \cellcolor{gray!20} Korak et al.~\cite{korak} & 2014 & &  & &  &  & Y & \makecell{N/A$^{\dag}$} & \makecell{\emph{Bypass security verification checks.}\\\emph{AES key recovery}}\\\cline{1-3}\cline{5-5}\cline{8-10}

 \cellcolor{gray!20} Dehbaoui et al.~\cite{dehbaoui2013electromagnetic}  & 2013 & \multirow{12}{*}{EMFI} &   & \multirowcell{2}[-3pt]{ARM Cortex-M3}  & &  & Y & \makecell{Two correct and two faulty\\plaintext-ciphertext pairs} & \makecell{AES key recovery.}\\\cline{1-2} \cline{8-10}
  
 \cellcolor{gray!20} Moro et al.~\cite{moro2013electromagnetic}  & 2013 &  &  & & &  & Y & \makecell{N/A$^{\dag}$} & \makecell{Load unauthorised data into RAM.\\Bypass verification checks.}\\\cline{1-2} \cline{5-5}\cline{8-10}
   
 \cellcolor{gray!20} Riviere et al.~\cite{riviere2015high} & 2015 &  &  &\makecell{ARM Cortex-M4} & &  & Y &  \makecell{Up to 96\%} & \makecell{\emph{AES and RSA-CRT key recovery}.$^{\dag}$\\\emph{Privilege escalation}.$^{\dag}$}\\\cline{1-2} \cline{4-10}
    
BADFET~\cite{cui2017badfet} & 2017 &  & Single-core & \makecell{Cisco 8861 IP Phone\\Broadcom BCM11123 SoC} & \makecell{Exploitable TrustZone SMC code} & OEM & N & \makecell{72\%} & \makecell{Bypass secure boot verification.\\Load unauthorised bootloaders.\\Privileged TEE access.}\\\cline{1-2} \cline{4-10}
     
\cellcolor{gray!20} Liao \& Gebotys~\cite{liao2019methodology}  & 2019 & & \multirowcell{5}[-3pt]{MCU} & \makecell{Microchip PIC16F687} & \makecell{---} & Reference & Y  & \makecell{Up to 222 EM pulses and\\5.3 plaintexts required on average} & \makecell{AES key recovery.\\\emph{Bypass secure boot verification.}$^{\dag}$}\\\cline{1-2} \cline{5-10}
      
\cellcolor{gray!20} Menu et al.~\cite{menu2019precise}  & 2019 &  &  & \makecell{Atmel SAM3X8E\\ARM Cortex-M3} & \makecell{AES key resides in Flash memory.} & OSS & Y & \makecell{Up to 100\%} & \makecell{AES key recovery.\\AES key resetting.}\\\cline{1-2} \cline{5-10}

\cellcolor{gray!20} Elmohr et al.~\cite{elmohr}  & 2020 &  &  & \makecell{NXP LPC1114\\ARM Cortex-M0\\E31 RISC-V\\SiFive FE310-G002} & \makecell{NXP LPC1114 chip decapsulation} & Reference & Y & \makecell{12\%--31\%} & \makecell{\emph{Bypass verification checks}.$^{\dag}$.\\Load unauthorised data into RAM.}\\\cline{1-2} \cline{4-10}

Gaine et al.~\cite{gaine2020emfi} & 2020 & & \makecell{Multi-core\\1GHz+}  & ARM Cortex-A53 & --- & \makecell{Linux kernel\\(\texttt{libpam})} & Y & 0.35\%--2\% (DVFS on--off) & Linux privileged access (REE).\\\hline%

\end{tabular}

    \begin{tablenotes}
    \item \textbf{Gray-coloured works do not target mobile SoCs.}
\item *: Device model and manufacturer are not disclosed, \dag: Discussed as a potential attack but not experimentally verified, Y: Yes, N: No.
\item \textbf{OEM}: OEM implementation, \textbf{OSS}: Open-source software, \textbf{Reference}: Reference implementation.
\item \textbf{FI}: Fault injection, \textbf{LFI}: Laser-based FI, \textbf{VFI}: Voltage-based FI, \textbf{CFI}: Clock-based FI, \textbf{HFI}: Heating-based FI, \textbf{EMFI}: Electromagnetic FI, \textbf{SW}: Software, \textbf{PMIC}: Power management integrated circuit, \textbf{MCU}: Microcontroller unit, \textbf{PBC}: Pairing-based cryptography, \textbf{PLL}: Phase locked loop, \textbf{SoC}: System-on-chip, \textbf{DVFS}: Dynamic voltage and frequency scaling.
\end{tablenotes}
\end{threeparttable}
\end{adjustbox}
\end{sidewaystable*}

\begin{sidewaystable*}
\begin{adjustbox}{width=\columnwidth}
\begin{threeparttable}
\caption{Summary of recent physical side-channel attacks on mobile and embedded systems.}
\label{tab:comparative-sca}
\def\arraystretch{1.5}
\begin{tabular}{|c|c|c|c|c|c|c|c|c|c|}
\hline
\rowcolor{black!85}
 \textcolor{white}{\textbf{Work}} &  \textcolor{white}{\textbf{Year}} & \textcolor{white}{\textbf{Class}} & \textcolor{white}{\textbf{SoC Type}} & \textcolor{white}{\textbf{Evaluation Platforms}} & \textcolor{white}{\textbf{Prerequisites}} & \textcolor{white}{\textbf{Implementation}} & \textcolor{white}{\textbf{Trigger?}} & \textcolor{white}{\textbf{Success Rate/Criteria}} &  \textcolor{white}{\textbf{Attack Goals}} \\\hline
 %
 %
Aboulkassimi et al.~\cite{aboulkassimi2011electromagnetic} & 2011 & \multirow{25}{*}{EM-SCA} & \multirowcell{2}[-5pt]{Single-core} & \makecell{Java ME phone*\\32-bit RISC CPU (350MHz)}  & MicroSD interface & \makecell{Bouncy Castle \\and Reference} & Y & \makecell{250 traces} & AES key recovery.\\\cline{1-2} \cline{5-10}

Kenworthy \& Rohatgi~\cite{kenworthy2012mobile}  & 2012 & & & \makecell{4G LTE smartphone*\\Mobile PDA*}  & \makecell{Square-and-multiply RSA\\Double-and-sometimes-add ECC} & \makecell{OEM, OSS and\\Reference} & ? & \makecell{Single trace (RSA and ECC)\\12,500 AES block op. traces} & \makecell{RSA, ECC and AES key recovery.}\\\cline{1-2} \cline{4-10}

\cellcolor{gray!20} Montminy et al.~\cite{montminy2013differential} & 2013 & & MCU &\makecell{Stellaris LM4F232H5QD\\ARM Cortex-M4F (50MHz)}  &---& OEM & N & \makecell{100,000 traces} & \makecell{AES key recovery.}\\\cline{1-2} \cline{4-10}

Nakano et al.~\cite{nakano2014pre} & 2014 &  & Single-core & \makecell{Android smartphone* (832MHz)}  & Square-and-multiply RSA  & Android JCE & N & \makecell{Single trace} & \makecell{ECC and RSA key recovery.}\\\cline{1-2}\cline{4-10}

Balasch et al.~\cite{balasch2015dpa} & 2015 &  & \makecell{Single-core\\1GHz+} & \makecell{ARM Cortex-A8 (1GHz)\\TI AM3358 Sitara SoC\\Beaglebone Black}  & --- & Reference & Y & \makecell{400,000 traces (masked AES)\\10,000 traces (unmasked AES)} & Masked AES key recovery.\\\cline{1-2} \cline{4-10}

Goller \& Sigl~\cite{goller2015side} & 2015 & & ? & \makecell{Four smartphones*\\\emph{Samsung Galaxy S3} (see text)}  & \makecell{Square-and-multiply RSA\\External casing is opened}  & Android JCE & N & \makecell{276 traces (EM shielding)\\170 traces (no EM shielding)} & RSA and AES key recovery. \\\cline{1-2}\cline{4-10}

Longo et al.~\cite{longo2015soc} & 2015 & & \multirowcell{4}[-2pt]{Single-core\\1GHz+} & \makecell{ARM Cortex-A8 (1GHz)\\TI AM3358 Sitara SoC\\Beaglebone Black SBC}  & \makecell{Cryptographic co-processor} & \makecell{OEM and OpenSSL\\(v1.0.1j)}  & Y & \makecell{20,000 traces (SW AES)\\500M traces (co-processor AES)} & \makecell{AES key recovery.\\Co-processor AES key recovery.\\\emph{FDE key recovery.}$^{\dag}$}\\\cline{1-2}\cline{5-10}

Belgarric et al.~\cite{belgarric2016side} & 2016 &  &   & \makecell{Android smartphone*\\Qualcomm MSM7225 SoC}  & External casing is opened &  \makecell{Bouncy Castle\\(v1.5)} & Y & 39 traces & \makecell{ECDSA key recovery.}\\\cline{1-2} \cline{5-10}

Genkin et al.~\cite{genkin2016ecdsa} & 2016 & &   & \makecell{Sony-Ericsson Xperia X10\\Apple iPhone 3GS}  & --- & \makecell{OpenSSL (v1.0.x)\\iOS 7.1.2–8.3} & N & 5000 traces & ECDSA key recovery.\\\cline{1-2} \cline{4-10}
 
\cellcolor{gray!20}
 Saab et al.~\cite{saab2016side} & 2016 && \makecell{Multi-core\\1GHz+} & \makecell{Intel i7-3517UE}  & Intel Ivy Core architecture& \makecell{Intel AES-NI\\(Ivy Bridge)} & N & \makecell{1.3--1.5 million traces} & AES key recovery.\\\cline{1-2} \cline{4-10}

Bukasa et al.~\cite{bukasa2017trustzone}& 2017 &  & Multi-core & \makecell{Raspberry Pi 2\\Broadcom BCM2836 SoC\\ARM Cortex-A7 (900MHz)}  & --- & Reference & Y & \makecell{17.81\%--38.30\% with 150,000 traces} & Secure world AES key recovery.\\\cline{1-2} \cline{4-10}
 
\cellcolor{gray!20} Camurati et al.~\cite{camurati2018screaming} & 2018 & & \makecell{Multi-signal\\IoT SoC} & \makecell{Nordic nRF52832\\ARM Cortex-M4\\Qualcomm Atheros AR9271}  & Mixed-signal SoC &  \makecell{\texttt{tinyAES} and\\\texttt{mbedTLS}\pmark} & N & \makecell{70,000 traces (\texttt{tinyAES})\\40,000 traces (\texttt{mbedTLS})} & AES key recovery.\\\cline{1-2} \cline{4-10}

Alam et al.~\cite{alam2018one} & 2018 & & Single-core & \makecell{Samsung Galaxy Centura\\Alcatel Ideal}  & \multirowcell{2}[-3pt]{---} & OpenSSL (v1.1.0g) & N &\makecell{Single trace (95.7\%--99.6\% recovery)} & \makecell{Constant-time\\RSA key recovery.}\\\cline{1-2} \cline{4-5}\cline{7-10}

Leignac et al.~\cite{leignac2019comparison} & 2019 & & \makecell{Multi-core\\1GHz+} & \makecell{HiKey LeMaker SBC\\ARM Cortex-A53 (1.2GHz)}  &  & Undisclosed & Y &\makecell{6000 traces} & REE and TEE AES key recovery.\\\cline{1-2} \cline{4-10}

Vasselle et al.~\cite{vasselle} & 2019 & & ? & \makecell{Undisclosed}  & Decomposable PoP SoC & Undisclosed & N &\makecell{2500 traces, two ciphertexts} & \makecell{Co-processor AES key recovery.\\Boot-time firmware decryption.}\\\cline{1-2} \cline{4-10}

 \cellcolor{gray!20} Wang et al.~\cite{wang2020far} & 2020 &  & IoT SoC & \makecell{Nordic nRF52832 SoC}  & --- & \texttt{tinyAES}\pmark & Y & \makecell{367 traces} & \makecell{AES key recovery.}\\\cline{1-5} \cline{6-10}
 
Lisovets et al.~\cite{lisovets2021iphone} & 2021 & \makecell{EM-SCA+\\P-SCA} & \makecell{Multi-core\\1GHz+} & \makecell{Apple iPhone 4\\Apple A4 SoC}  & \makecell{Exploitable boot ROM\\Unlimited AES engine encryptions}& OEM & Y & \makecell{300M traces (EM-SCA)\\200M traces (P-SCA)} & \makecell{AES key recovery.}\\\cline{1-5} \cline{6-10}


\cellcolor{gray!20} Schmidt et al.~\cite{schmidt2010side} & 2010 & \multirow{10}{*}{P-SCA} & \multirow{8}{*}{MCU} & \makecell{NXP LP2148 (ARM7TDMI-S)\\AVR ATMega163 + AT89S2853}  & Exposed I/O pins & \multirow{8}{*}{Reference} & Y &\makecell{N/A$^{\dag}$} & Key recovery.\\\cline{1-2}\cline{5-6} \cline{8-10} 

\cellcolor{gray!20} Heuser \& Zohner~\cite{heuser2012intelligent} & 2012 & &  & \makecell{AVR ATMega-256-1}  & \multirow{7}{*}{Accessible power line} &  & Y &\makecell{20--60 traces (low--high noise)} & \makecell{AES key recovery.}\\\cline{1-2} \cline{5-5} \cline{8-10} 
 
\cellcolor{gray!20} Msgna et al.~\cite{msgna2014precise} & 2014 & &  & \makecell{AVR ATMega163}  & &  & N & \makecell{66.78--100\% (11 instructions)} & \makecell{Reverse engineering.}\\\cline{1-2} \cline{5-5} \cline{8-10} 

\cellcolor{gray!20} Maghrebi et al.~\cite{maghrebi2016breaking} & 2016 & &  &  \multirow{3}{*}{AVR ATMega328P}   &   & & N & \makecell{20 traces (unprotected AES)\\500 traces (masked AES)} & Masked AES key recovery.\\\cline{1-2}  \cline{8-10}

\cellcolor{gray!20} Picek et al.~\cite{picek2018performance} & 2018 & &  & &   & & N & \makecell{10 traces} & AES key recovery.\\\cline{1-2}  \cline{8-10} 

\cellcolor{gray!20} Park et al.~\cite{park2018power} & 2018 & &  &  &   &  & Y & \makecell{99.03\% (112 instructions)} & \makecell{Reverse engineering.}\\\cline{1-2} \cline{5-5} \cline{8-10}

\cellcolor{gray!20}  Wang \cite{wang2019side} & 2019 &  &  &AVR ATXmega128D4  &   &  & N & 160.3--400.2 traces & AES key recovery.\\\cline{1-2}\cline{4-10}

\cellcolor{gray!20} Gnad et al.~\cite{gnad2019leaky} & 2019 &  & \makecell{Mixed-signal\\IoT SoC} &\makecell{STM 32L475 and 32F407VG \\ESP32-devkitC}  &  Mixed-signal SoC & \makecell{\texttt{tinyAES} and\\\texttt{mbedTLS}\pmark} & Y & \makecell{10M traces} & AES key recovery.\\\hline

\cellcolor{gray!20} Genkin et al.~\cite{genkin2014rsa} & 2014 & AC-SCA & \makecell{Multi-core\\1GHz+} & Lenovo laptop  &  --- & GnuPG (v1.x)  & N & One-hour audio trace & RSA key recovery.\\\hline

\end{tabular}

\begin{tablenotes}
\item \textbf{Gray-coloured works do not target mobile SoCs.}
\item ?: Method not disclosed. *: Device model and manufacturer are not disclosed, \dag: Discussed as a potential attack but not experimentally verified, Y: Yes, N: No.
\item \textbf{OEM}: Proprietary OEM implementation, \textbf{OSS}: Open-source software implementation, \textbf{Reference}: Reference implementation, \pmark: Implementation and/or version not disclosed.
\item \textbf{SCA}: Side-channel attack, \textbf{EM-SCA}: Electromagnetic SCA, \textbf{P-SCA}: Power-analysis SCA, \textbf{AC-SCA}: Acoustic SCA, \textbf{JVM}: Java virtual machine, \textbf{Java ME}: Java Platform -- Micro Edition, \textbf{JCE}: Java cryptography extensions library, \textbf{ECC}: Elliptic curve cryptography, \textbf{SBC}: Single-board computer, \textbf{SoC}: System-on-chip, \textbf{FDE}: Full-disk encryption.
\end{tablenotes}
\end{threeparttable}
\end{adjustbox}
\end{sidewaystable*}

\subsection{Challenges and Limitations}
\label{sec:challenges}

Applying fault injection and side-channel attacks on mobile devices poses unique difficulties for researchers compared to traditional computing systems. Among the principal challenges, digital subsystems---CPUs, GPUs, DRAM, flash memory, etc.---are connected at significant distances ($>$10cm) over relatively accessible buses on traditional systems. In contrast, these components are condensed into one or more millimeter-sized packages using wiring distances measured at the micrometer scale on modern mobile SoCs. This renders wire-level or pin-level attacks extremely difficult on mobile devices without risking permanent damage to the target.  To compound this issue, the density of mobile SoCs, the use of high-frequency CPUs, and mixed-signal components generate acute levels of electrical interference that must be disentangled to accurately target operations of interest~\cite{aboulkassimi2011electromagnetic,genkin2016ecdsa,nakano2014pre}.

Compare this to existing work against desktop computers: VoltPillager~\cite{voltpillager} subverted Intel SGX enclaves by targeting an accessible voltage regulation interface on a desktop motherboard, while Govindavajhala and Appel~\cite{govindavajhala2003using} used a 50W placed near visible DRAM chips for heating-assisted fault injections. Such attacks leverage an important advantage of targeting desktop computers: component sizes and wiring distances are, on the whole, significantly larger and can be targeted with a greater degree of precision using lower cost equipment. While attacks against traditional computing systems share some similarities to mobile platforms---for example, the shared use of statistical techniques like differential fault and power analysis---differences in their physical form factors introduce major technical challenges that must be overcome. 

Further to these inherent limitations, we have also identified some methodological shortcomings in the current state of the art. The rest of this section discusses these issues and provides recommendations to researchers in the field for their remediation. Furthermore, we pose several open research challenges arising from some major difficulties for executing FIAs and SCAs on mobile devices.

\subsubsection{Few Attacks Against Implementations with Known Countermeasures}
From the inception of physical fault injection and side-channel attacks, a range of hardware- and software-based countermeasures have been developed to mitigate their effectiveness. Specifically, hardware-based power filters have been known for 20 years to thwart voltage glitches~\cite{kim2007faults}; EM shielding units, such as Faraday cage packages~\cite{quisquater2001electromagnetic}, can mitigate EMFIs and EM SCAs; and frequency detection circuits and hardware frequency locking can be deployed to hinder clock glitches. 
Hardware- and software-based redundancy checks of security-critical procedures and introducing randomness into program execution, e.g.\ random timing waits and CPU clock frequency fluctuations, have also been touted as general solutions~\cite{bar2006sorcerer}.

While these methods are known in the community, recent research has still focussed on evaluating unprotected reference implementations. Consequently, the reported success rates of many attacks most likely represent idealistic testing scenarios. In some work, researchers have used implementations that are known to exhibit side-channel leakage, such as double-and-sometimes-add ECC and square-and-multiply RSA~\cite{kenworthy2012mobile}, in place of constant-time, blinded, masked and fixed-window counterparts. \textbf{We observed generally that cryptographic co-processors have been subjected to extremely few attacks}~\cite{longo2015soc,vasselle2019breaking}. \textbf{Indeed, there are no published attacks against state-of-the-art OEM co-processors with known countermeasures}, e.g.\ Apple's SEP (\S\ref{sec:proprietary}) and Google's Titan M (\S\ref{sec:google_titan}).  Protected implementations are not invulnerable, but they usually increase the requirements, e.g.\ number of traces, by some orders of magnitude~\cite{balasch2015dpa,maghrebi2016breaking,longo2015soc}.

\begin{itemize}
    \item[$\bigstar$] \textbf{Recommendation 1}: We recommend that researchers investigate implementations with established countermeasures as part of an evaluation test-bed and provide appropriate justification where this is not possible.
\end{itemize}

\subsubsection{Generalisability and Transparency Issues}

Published research has also tended to focus on a small set of evaluation platforms. Specifically, \textbf{only a single target was investigated in 80\%+ fault injection and 60\%+ side-channel attack publications examined in this survey}. The reported success rates of single-target research papers are vulnerable to device-specific biases and may, therefore, be a significant overestimate when applying them to alternative targets. SoC-by-SoC differences in power management systems, voltage regulators, power planes, land-side decoupling capacitors (LDCs), clock frequencies and CPU architectures may pose hidden issues that severely hinder if not prevent the reproducibility of such attacks. 

Poor methodological transparency is also an issue: \textbf{we note that many authors simply do not disclose proof-of-concept code or even the target under test}~\cite{aboulkassimi2011electromagnetic,goller2015side,kenworthy2012mobile,nakano2014pre,ait2019analyzing,vasselle2019breaking}. Only high-level details are shared in some cases, e.g.\ \emph{``an Android smartphone''}~\cite{nakano2014pre} or a \emph{``4G LTE smart phone from a major manufacturer''}~\cite{kenworthy2012mobile}, while, in other work, this information is clearly redacted~\cite{ait2019analyzing,vasselle2019breaking}.  In addition, Benadjila et al.~\cite{benadjila2020deep} criticised the closed-source nature of SCA data analysis frameworks: \emph{``hyperparameterisation }[of machine learning models] \emph{has often been kept secret by the authors who only discussed the main design principles and on the attack efficiencies.''} Critically, Wang~\cite{wang2019side} showed that the effectiveness of such models can vary severely between devices: a deep learning model with a purported 96\% accuracy dropped to only 2.45\% when tested on traces captured from another board with the same IC. 

\begin{itemize}
    \item[$\bigstar$] \textbf{Recommendation 2}: In pursuit of open science, we strongly advise researchers publish proof-of-concept code, precise device/SoC models and any equipment details needed to reproduce attacks wherever possible. (To this end, we welcome the move by prominent events to normalise the submission of paper artifacts~\cite{ches2021artifacts,eurosp2021artifacts,usenix2021artifacts}).
\end{itemize}

\subsubsection{Few Attacks Against Package-on-Package ICs}

Package-on-package designs pose new challenges as researchers must account for the target SoC's activity in addition to side-effects from other packages, e.g.\ DRAM modules. Alas, \textbf{we are not aware of any successful fault injection attacks on PoP systems used by modern mobile devices}, e.g.\ Apple's A14 on the iPhone 12 series.   A{\"\i}t El Mehdi \cite{ait2019analyzing} presented initial signs that PoP EMFIs are possible, but a complete attack could not be mounted.  To our knowledge, only two SCAs have been published against PoP SoCs. Vasselle et al.~\cite{vasselle2019breaking} used EM analysis for key recovery from an on-SoC cryptographic co-processor, but this exploited a unique part of the device's system initialisation process where executing a partially disassembled PoP was possible. 
Recent work by Lisovets et al.~\cite{lisovets2021iphone} demonstrated EM- and power-based SCAs against an Apple A4 PoP SoC on an iPhone 4. While there are complexities in generalising the attack to current Apple devices, the authors showed that PoP side-channel attacks are a practical reality.

\begin{itemize}
    \item[$\diamondsuit$] \textbf{Research Challenge 1}: We suggest greater focus on developing novel methods against PoP modules that are increasingly being used by mobile device vendors. We believe that this area presents pressing research challenges for overcoming interference generated by multiple ICs in a single package, particularly for fault injections where there are no publicly known attacks.
\end{itemize}

To assist in this area, we point to recent research that demonstrated the possibility of desoldering PoP components without alteration~\cite{heckmann2018forensic}. Combining conductive and insulating adhesives to stabilise PoP systems is a solution that can potentially be used for enabling fault injection attacks on PoP SoCs~\cite{felba2011thermally}. In digital forensics, polymeric adhesives are of particular interest that can be used for complex repairs or even in realising advanced man-in-the-middle attacks to reverse engineer secure systems. Heckmann et al.~\cite{heckmann2017electrically} (2017) illustrated this in a home-made memory man-in-the-middle attack platform. This platform aims to provide read/write access to data exchanged between the CPU and its volatile and non-volatile memory, providing a first step for fault injection attacks on PoP SoCs.

\subsubsection{Chosen-Plaintext and -Ciphertext Attack Feasibility}

A significant practical challenge of chosen-plaintext and -ciphertext attacks is the difficulty of collecting an adequate number of useful measurement traces and faulted outputs. In particular, voltage- and clock-based FIAs typically generate exploitable faults at an extremely low rate ($<$0.1\% of all faults)~\cite{controlling_pc_fi,timmers2017escalating,blomer2014practical}, while reported key recovery attacks have required 1--500 million traces or faulted ciphertexts from modern SoCs~\cite{balasch2015dpa,longo2015soc,vasselle2019breaking,saab2016side,lisovets2021iphone,gnad2019leaky}. Related to this issue, some work has required cryptographic operations to be performed multiple times under the same message and key~\cite{dehbaoui2013electromagnetic} and TEE applications to be invoked repeatedly without limitation~\cite{tang2017clkscrew,qiu2019voltjockey}. Depending on the target system, these conditions can be time-consuming to reproduce in practice. As a concrete example, targeting secure boot cryptographic operations, e.g.\ digital signature checks, may only be possible once per component and per power cycle. Data-demanding approaches may be useful if their conditions can be easily repeated and automated, otherwise they are unlikely to be feasible in time-constrained environments. 

\begin{itemize}
    \item[$\diamondsuit$] \textbf{Research Challenge 2}: Can the requirements of data-heavy attacks be minimised while remaining transferable across multiple targets?  As a potential way forward, transfer learning~\cite{weiss2016survey,thapar2020transca,hoang2020plaintext} has been recently explored for training a \emph{base} model using a large set of easily collectable measurements, which is then fine-tuned to a target device using a smaller dataset for capturing any platform-specific idiosyncrasies. This research avenue remains in its infancy, however.
\end{itemize}

\subsubsection{Inaccessible Hardware}

Accessing hardware components to inject faults and acquire exploitable measurements remains a major challenge on today's mobile devices. For power-based SCAs, voltage supply pins are not easily accessible on modern SoCs as they are occluded by ball-grid array (BGA) assembled packages; BGA-level analysis requires specialist test equipment and expertise, significantly increasing attack complexity~\cite{balasch2015dpa}. Indeed, invasive methods are needed just to evaluate the feasibility of an attack, let alone execute one successfully.   It is likely, for this reason, that researchers have used power-based SCAs and FIAs attacks on platforms with easily accessible supply pins and power planes, e.g.\ SBCs and MCUs.  We also posit that this is behind the popularity of EM SCAs on mobile SoCs because of their predominantly non-invasive means of signal acquisition.

\begin{itemize}
    \item[$\diamondsuit$] \textbf{Research Challenge 3}: Moving beyond the current state of play, can low-cost methods be developed for interrogating BGA-assembled PoP/SoC packages in order to modernise power-based side-channel and fault injection attacks?
\end{itemize}

A related issue is the lack of publicly available data sheets provided by OEMs and SoC vendors. Unfortunately, without a radical shift to open hardware, this is difficult to circumnavigate without movement from device and silicon vendors. This has forced researchers into evaluating devices under test using black-box methods, raising a security by obscurity concern: the absence of attacks on a given SoC does not imply that it is secure against SCAs and FIAs. Given the large number of commercially available SoCs, fault injections and/or side-channel attacks may be possible if previously unevaluated SoCs undergo a dedicated investigation.

\subsubsection{Risk of Device Damage}
\label{sec:device_damage}
Fault injections, by definition, leverage unexpected behaviour when the target is subjected to conditions beyond its intended operating conditions. Over-glitching, e.g.\ extreme over- and under-volting, can inflict permanent and unpredictable damage on device components, often in ways that are not immediately observable. In practice, this is likely to occur during the exploration of previously unknown glitch parameters. This extends to recent side-channel attacks requiring the decapsulation of ICs and disassembling of PoPs to enhance the acquisition of exploitable traces. In some work, preparation techniques have been refined on spare components before mounting the actual attack~\cite{vasselle2019breaking}. Researchers operating on a budget or with access to only a single or a small number of relevant SoCs may struggle to replicate such attacks.

In digital forensics, investigators have also resorted to laser or chemical decapsulation to gain appropriate hardware access for reverse engineering and/or data extraction~\cite{staller2010low}. Such methods significantly increase the risk of destroying the target IC, particularly its wire bonding~\cite{heckmann2019decrease}. This damage can occur in two ways: 1), from the direct application of high-energy lasers or corrosive chemicals~\cite{heckmann2019decrease}; and 2), when manipulating probes during chip-on SCAs and FIAs or to monitor on-SoC memory units, buses and other peripherals~\cite{heckmann2017electrically}. As a risk mitigation measure, Staller~\cite{staller2010low} (2010) proposed a method coupling electrically conductive adhesives (ECAs) and the precision of laser ablation---also known as photoablation---for the repair of damaged bonding wires and to minimise permanent device damage.

\begin{itemize}
    \item[$\diamondsuit$] \textbf{Research Challenge 4}: Can safer interrogation techniques be developed for mobile SoCs and, in particular, PoP systems without necessitating elaborate preparation methods or the application of high-energy lasers and corrosive substances?
\end{itemize}

\subsection{Future Directions}
\label{sec:future}

The growth of mixed-signal and multi-SoC architectures are an important development for advancing the state of the art in fault injection and side-channel attacks on mobile systems. Recent research, e.g.\ \cite{camurati2018screaming} and \cite{gnad2019leaky}, showed that these can offer new vectors for recovering secret data. Given the economic incentives to miniaturise hardware into ever-more compact packages, it is likely that this trend will continue, and that new attack avenues may be opened from the interaction between increasingly diverse and closely located SoC components. In parallel, manufacturers continue to delegate long-running but low-complexity device features, such as sensor hubs, to separate device SoCs. This paradigm, also known as heterogeneous `big-little' architectures, permits the relatively power-hungry primary SoC to remain in a low-powered state while less energy-intensive SoCs are used for low-complexity processing. Similarly, novel interactions between these heterogeneous SoCs could yield novel attack methods in future research.

A second trend is the rise of deep learning-based approaches for side-channel analysis, which can model highly non-linear interactions for complex classification tasks.  This contrasts with traditional statistical methods that use carefully chosen parametric models. While this area is in its relative infancy, preliminary results demonstrate that fewer traces may be required for the same if not an improved level of effectiveness relative to correlation- and template-based analyses~\cite{hoang2020plaintext,picek2018performance,wang2019side}.

Researchers should also be aware of emerging mobile TEE applications in recent research. Examples include secure mobile deep learning~\cite{kunkel2019tensorscone}, protecting cryptocurrency wallets~\cite{gentilal2017trustzone}, authenticating adverts from mobile advertising networks~\cite{li2015adattester}, preserving the integrity of system logs~\cite{shepherd2017emlog,karande2017sgx}, novel remote attestation mechanisms~\cite{shepherd2017establishing,shepherd2018remote}, protecting healthcare data~\cite{segarra2019using}, and confidential image processing~\cite{brito2016arm}. Such proposals could serve as future attack targets. Moreover, new TEEs and security mechanisms continue to be developed using the RISC-V open-source instruction set architecture~\cite{lee2020keystone,menon2017shakti,schrammel2020donky,shepherd2021lira}. Examining the application of SCAs and FIAs against these systems will gain importance if RISC-V becomes widely adopted by mobile device OEMs.

Finally, we observe that manufacturers are increasingly returning to discrete secure element (SE) hardware for protecting the most critical device assets, such as cryptographic keys and user credentials, rather than TEEs. This is highlighted by Google's recent release of the Titan M module (\S\ref{sec:google_titan}) and the formation of the Android Ready SE Alliance for accelerating the adoption of SEs on Android devices~\cite{google:android_ready}. The alliance identifies that using tamper-resistant hardware for hosting digital keys for physical items, e.g.\ car keys; ePassport and national identity credentials; and digital wallets \emph{``offers the best path for introducing these new consumer use cases in Android''}~\cite{google:android_ready}. These targets have significant protections against FIAs and SCAs, which necessitates the development of fresh approaches.
\section{Conclusion}
\label{sec:conclusion}

Data extraction methods from mobile devices have significantly increased in difficulty due to the growing complexity of mobile SoCs and an increased focus on security. While this represents progress from a privacy perspective, it poses major challenges in the context of digital forensics, such as lawful evidence recovery. In this report, we presented an extensive survey of state-of-the-art physical fault injection and side-channel attacks on mobile system-on-chips. In total, over 50 research publications published between 2009 and 2021 were examined, which were individually mapped to their relevant attack goals, prerequisites, published success rates, and evaluated platforms.  Our aim was to consolidate the large base of existing literature into a format that can be digested by working practitioners in the field.

Beyond this, we presented a series of challenges and limitations arising from the current literature, identifying a series of recommendations and open research challenges pertaining to fault injection and side-channel attacks on mobile systems.  In particular, we raised the issue of attack reproducibility: redacted platform details, an extremely small number of evaluation devices, increasingly complex SoCs, and poor analysis transparency may all lead to generalisation issues on current and future targets. Moreover, we identified that the regular targeting of unprotected reference or custom software implementations is a cause for concern. Notably, many works do not evaluate software and hardware implementations with widely known FIA or SCA countermeasures. As a consequence, it is likely that researchers in the field will face significant difficulties when applying many attacks to other platforms, particularly OEM devices.

Lastly, we discussed potential future developments in the state of the art. Specifically, the exploitation of mixed-signal SoCs, novel interactions in `big-little' multi-SoC architectures, and the application of deep learning methods are all potentially fruitful areas of research. In addition, the relative absence of published attacks against PoP architectures and OEM cryptographic co-processors with known countermeasures is a pressing concern in light of their growing use by mobile vendors. These could obsolete many published attacks on mobile phone SoCs, which we see as an important research area going forward.

\section*{Acknowledgements}

\textit{Funding:} This work received funding from the European Union's Horizon 2020 research and innovation programme under grant agreement No. 883156 (EXFILES). The authors would like to thank the EXFILES WP5 project partners for comments and discussions around the topic of this work. 




\printcredits
\footnotesize
\bibliographystyle{abbrv}

\bibliography{cas-refs}

\bio{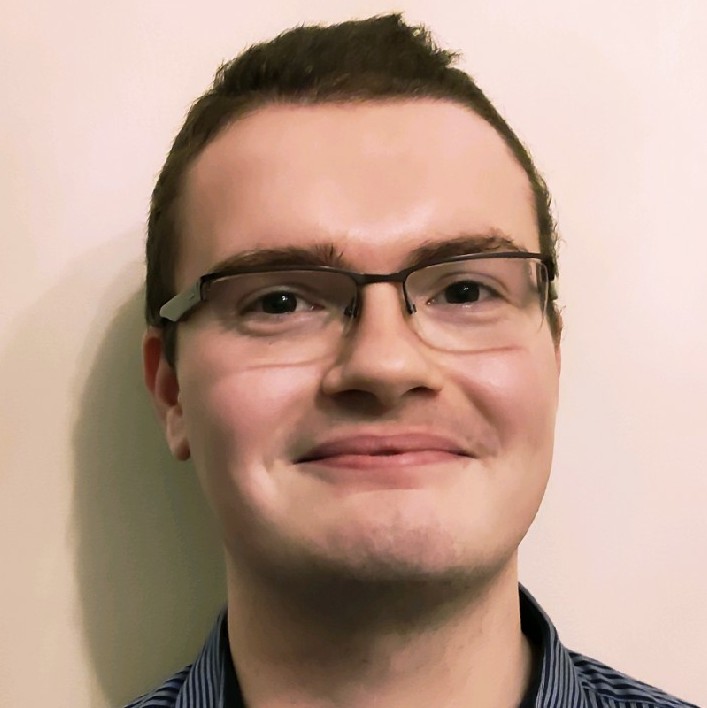}
\textbf{Carlton Shepherd} (B.Sc., Ph.D.) received his Ph.D. Information Security from the Information Security Group at Royal Holloway, University of London, and his B.Sc.\ in Computer Science from Newcastle University. He is currently a Senior Research Fellow at the Information Security Group at Royal Holloway, University of London. His research interests centre around the security of trusted execution environments (TEEs) and their applications, secure CPU design, embedded systems, applied cryptography, and hardware security.
\endbio

\bio{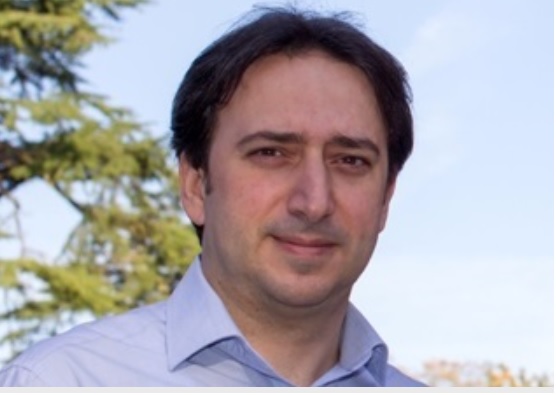}
\textbf{Konstantinos Markantonakis} (B.Sc., M.Sc., MBA, Ph.D.) is a Professor of Information Security in Royal Holloway University of London, and the Director of the Information Security Group Smart Card and IoT Security Centre (SCC). He obtained his B.Sc.\ (Lancaster University), M.Sc., Ph.D.\ (London) and his MBA in International Management from Royal Holloway, University of London. His research interests include smart card security and applications, secure cryptographic protocol design, embedded systems security, autonomous systems and trusted execution environments.
\endbio
\vspace{1.7cm}

\bio{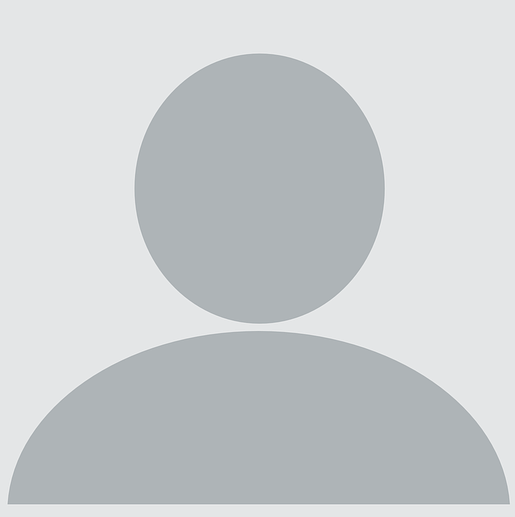}
\textbf{Nico van Heijningen} (M.Sc.) is a forensic researc-her at the Netherlands Forensic Institute (NFI). His main interest is applied cryptanalysis, which includes password recovery, fault injections and side-channel analysis.
\endbio
\vspace{0.7cm}

\bio{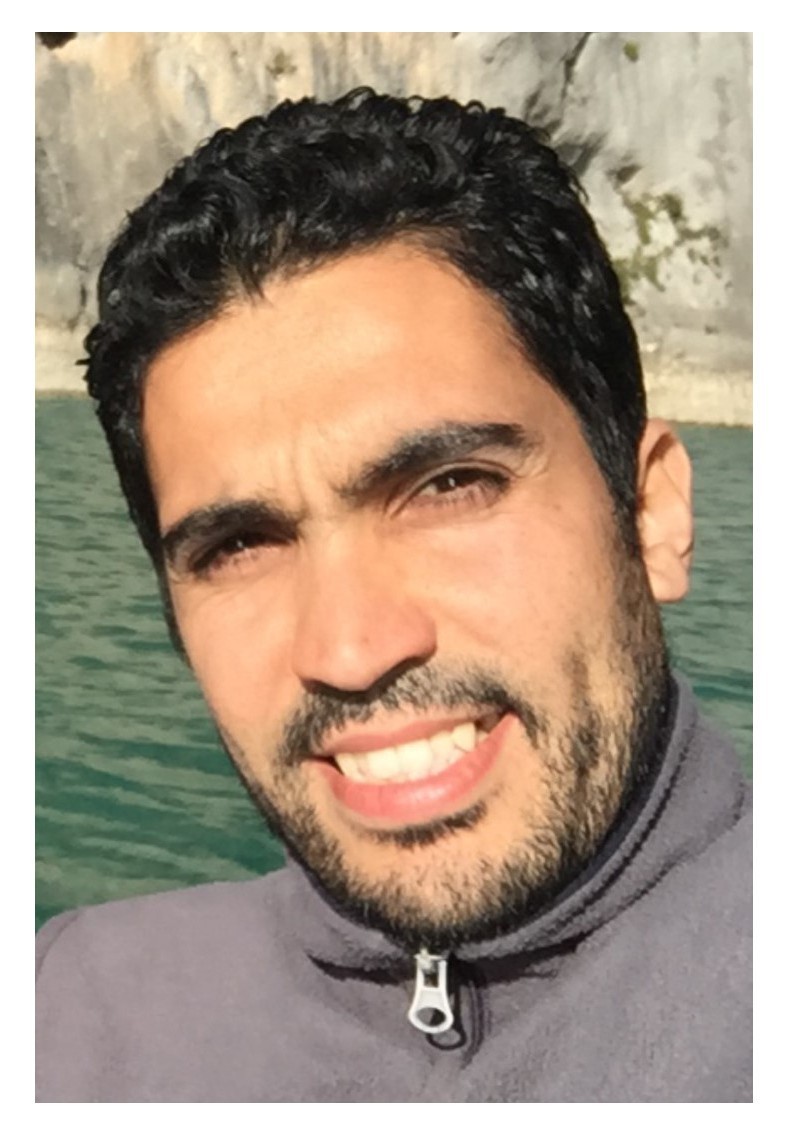}
\textbf{Driss Aboulkassimi} (S'09) received his master and engineering degree from the University of Montpellier. He joined the Secure Architectures and Systems (SAS) lab in 2010 from Ecole Nationale Superieure des Mines de St Etienne (ENSMSE) as a research engineer in the physical security characterisation of embedded systems. In 2014, Driss joined CEA-leti specialising in hardware security testing of mobile devices. He participates in and manages several European (EXFILES, MOBITRUST, HINT, PCAS) and French projects (CSAFE+, PACLIDO, TISPHANIE, BIOFENCE, SESAME, HOMERE, etc.).
\endbio

\bio{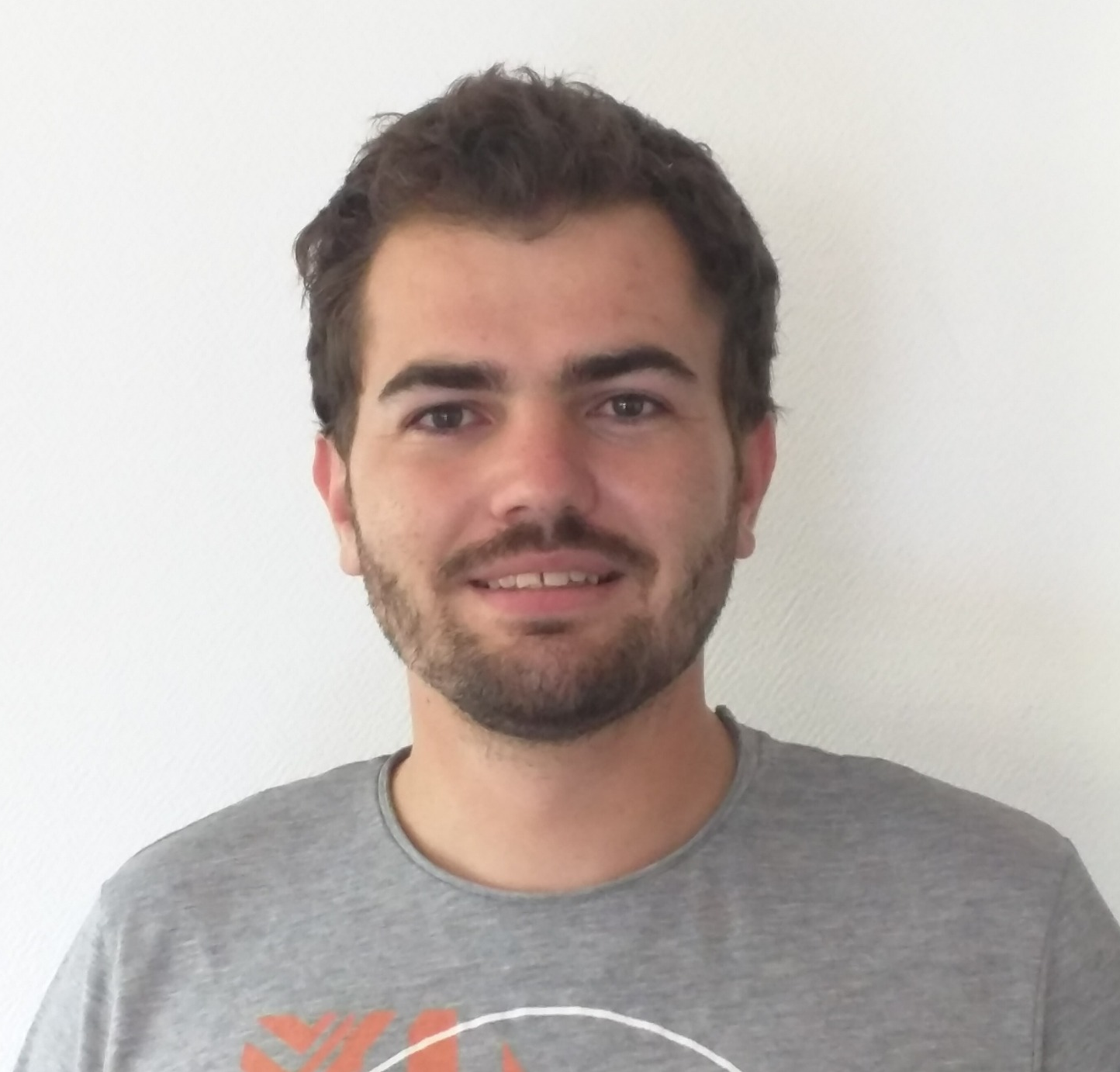}
\textbf{Cl\'{e}ment Gaine} is a Ph.D. student at the CEA in France. He is in the joint research team between the CEA and the Mines de Saint-Etienne, France. His research interests are fault injections and particularly electromagnetic fault injections.
\endbio
\vspace{0.7cm}

\bio{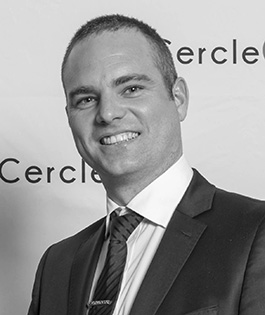}
\textbf{Thibaut Heckmann} (B.Sc., M.Sc., Ph.D.) is currently a military senior officer at the French national gendarmerie, associate researcher at the \'{E}cole normale sup\'{e}rieure of Paris, and a member of the national gendarmerie research center (CREOGN). From 2015 to 2020 he was the head of the national data extraction unit of the French gendarmerie Laboratory’s digital forensic department (IRCGN) and he was associate researcher at the Royal Holloway, University of London from 2017 to 2018. He received a M.Sc in fundamental physics and a Ph.D. in mathematics from the Ecole normale sup\'{e}rieure of Paris (ENS). In 2018, he received the European Emerging Forensic Scientist Award 2018-2021 from the European Academy of Forensic Science (EAFS) and the Cybersecurity Award from Cercle K2 in 2020.
\endbio

\bio{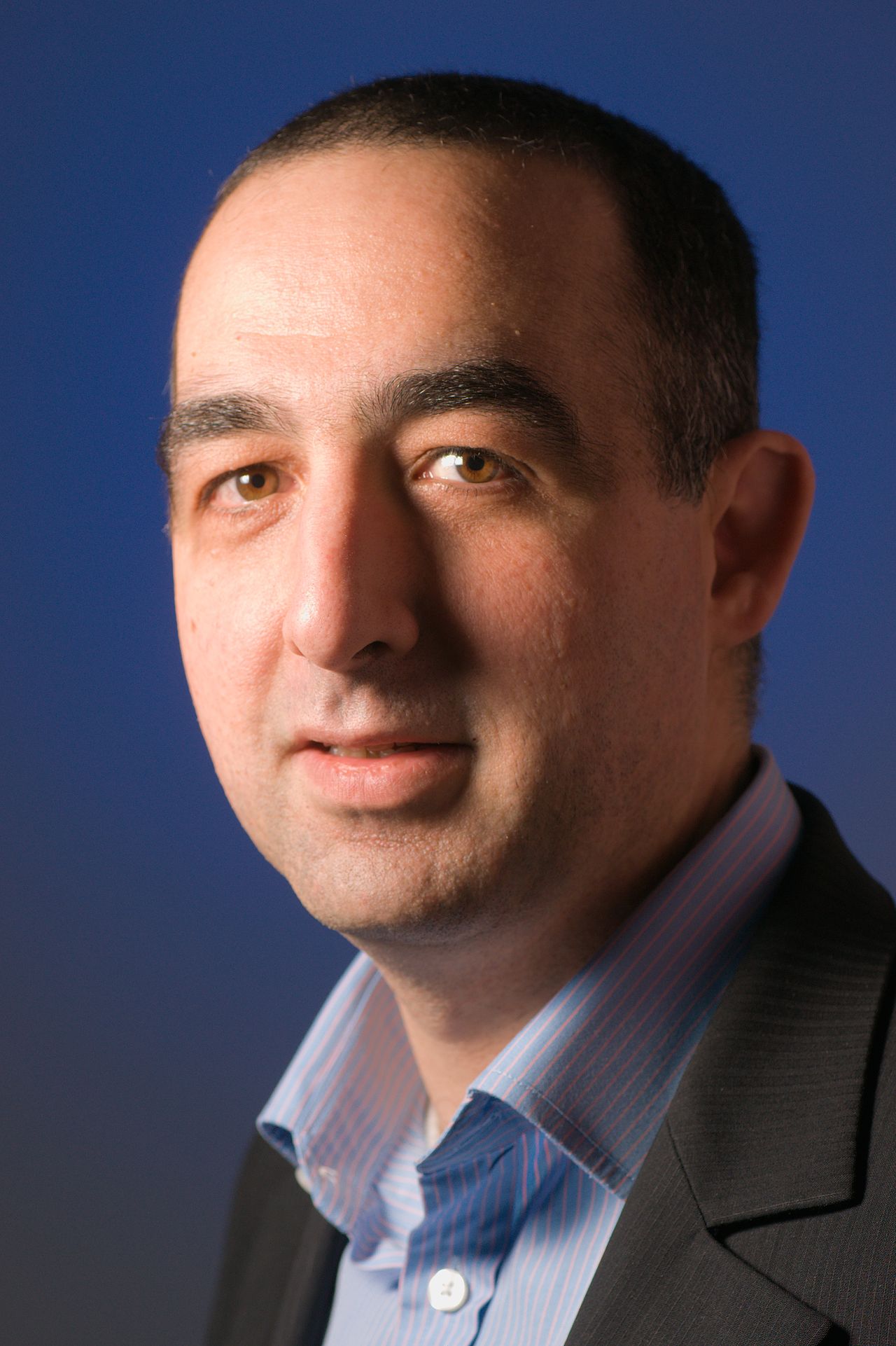}
\textbf{David Naccache} is a cryptographer, currently a professor at the \'{E}cole normale sup\'{e}rieure and a member of its Computer Laboratory. He was previously a professor at Panth\'{e}on-Assas University. He received his Ph.D. in 1995 from the \'{E}cole nationale sup\'{e}rieure des t\'{e}l\'{e}communications. His most notable work is in public-key cryptography, including the cryptanalysis of digital signature schemes. His areas of interest are side channel attacks, forensics and the detection of software flaws in embedded systems.
\endbio

\end{document}